\documentclass[preprint,12pt]{elsarticle}
\usepackage{amssymb}
\usepackage{amsmath}
\usepackage{CJKutf8}
\usepackage{multirow}
\usepackage{makecell}
\usepackage{graphicx}
\usepackage[table,x11names]{xcolor}
\usepackage{url}
\usepackage{hyperref}

\begin{document}

\begin{frontmatter}

\title{DGSNA: Dynamic Generative Scene-based Noise Addition method}

\author[label1]{Zihao Chen}
\ead{644913263@qq.com}
\author[label1]{Zhentao Lin}
\ead{2112005050@mail2.gdut.edu.cn}
\author[label1]{Bi Zeng}
\ead{zb9215@gdut.edu.cn}
\author[label2,label3]{Linyi Huang}
\ead{hly@ceprei.biz}
\author[label2,label3]{Jia Cai}
\ead{caijia@ceprei.biz}

\affiliation[label1]{organization={School of Computer Science and Technology, Guangdong University of Technology},
            addressline={No.100 Waihuan West Road},
            city={Guangzhou},
            postcode={510006},
            state={Guangdong},
            country={China}}
            
\affiliation[label2]{organization={China Electronic Product Reliability and Environmental Testing Research Institute},
            addressline={No.78 West Zhucun Avenue}, 
            city={Guangzhou},
            postcode={511370},
            state={Guangdong},
            country={China}}
            
\affiliation[label3]{organization={Key Laboratory of Ministry of Industry and Information Technology for Intelligent Products Testing and Reliability},
            addressline={No.78 West Zhucun Avenue}, 
            city={Guangzhou},
            postcode={511370},
            state={Guangdong},
            country={China}}

\begin{abstract}
To ensure the reliable operation of speech systems across diverse environments, noise addition methods have emerged as the standard solution.
However, existing methods offer limited coverage of real-world scenes and depend on pre-existing noise libraries and scene metadata.
This paper presents prompt-based Dynamic Generative Scene-based Noise Addition (DGSNA), a novel approach driven by generative language models that integrates Dynamic Generation of Scene-based Information (DGSI) with Scene-based Noise Addition for Speech (SNAS).
The DGSI module, with a BET (Background, Examples, Task) prompt framework, dynamically generates logic-compliant scene-based information, including scene dimensions, sound sources, and microphone positions, thereby addressing the challenges of scene enumeration and detailed description.
Complementing this, the SNAS module employs a Time-Frequency Diffusion-based (TFD) Text-to-Audio model to synthesize scene-specific noise. By integrating this noise with clean speech via Room Impulse Response (RIR) filters, the module streamlines the traditionally labor-intensive process of replicating diverse acoustic environments.
Experimental results show that DGSNA significantly enhances the robustness of speech recognition and keyword spotting models, achieving relative improvements of up to 11.32\%. Furthermore, DGSNA is highly compatible with existing noise addition techniques. Our implementation and demonstrations are available at https://dgsna.github.io.
\end{abstract}

\begin{graphicalabstract}
    \includegraphics[width=\textwidth]{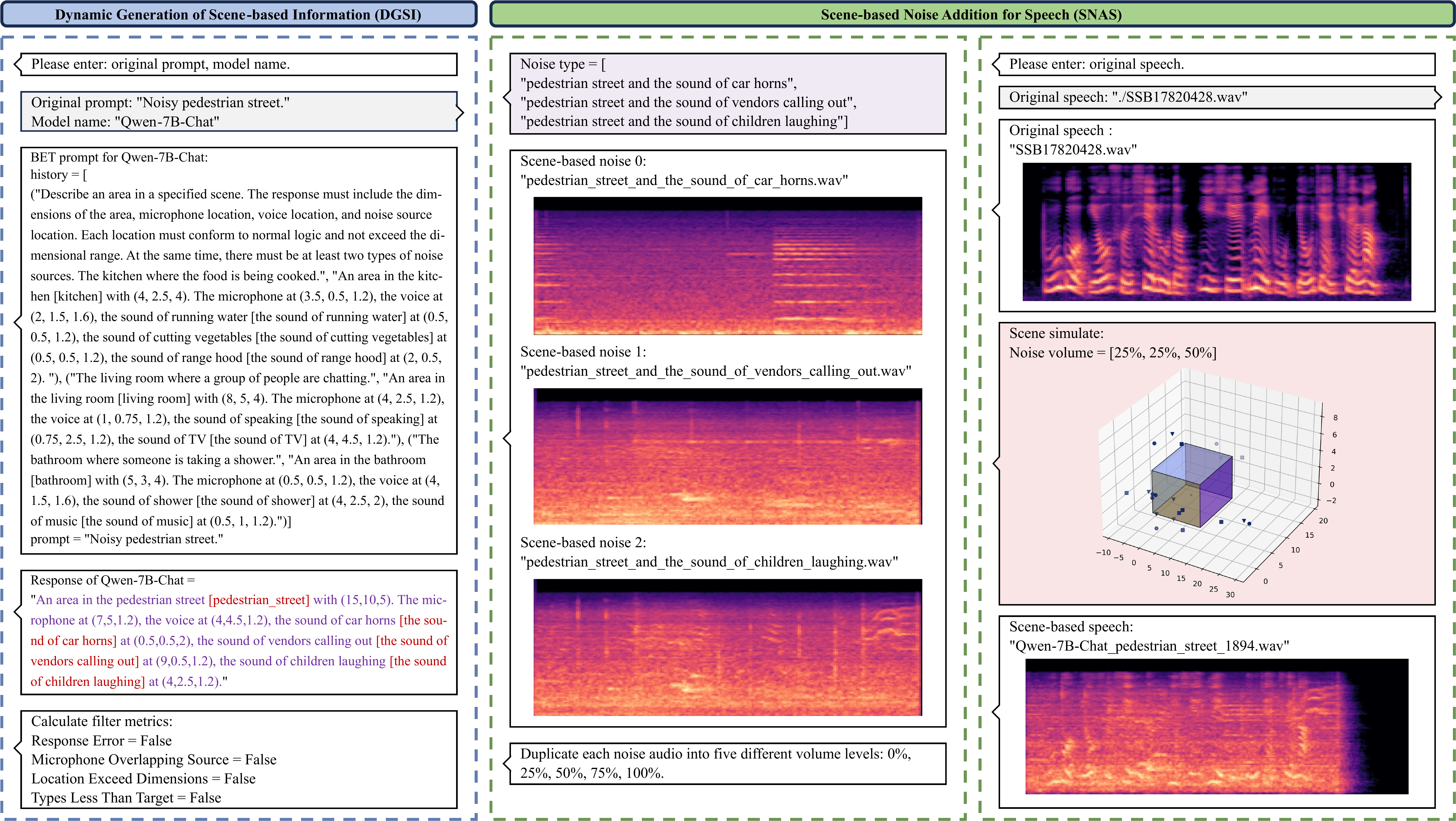}
    \vspace{10cm}
    \includegraphics[width=\textwidth]{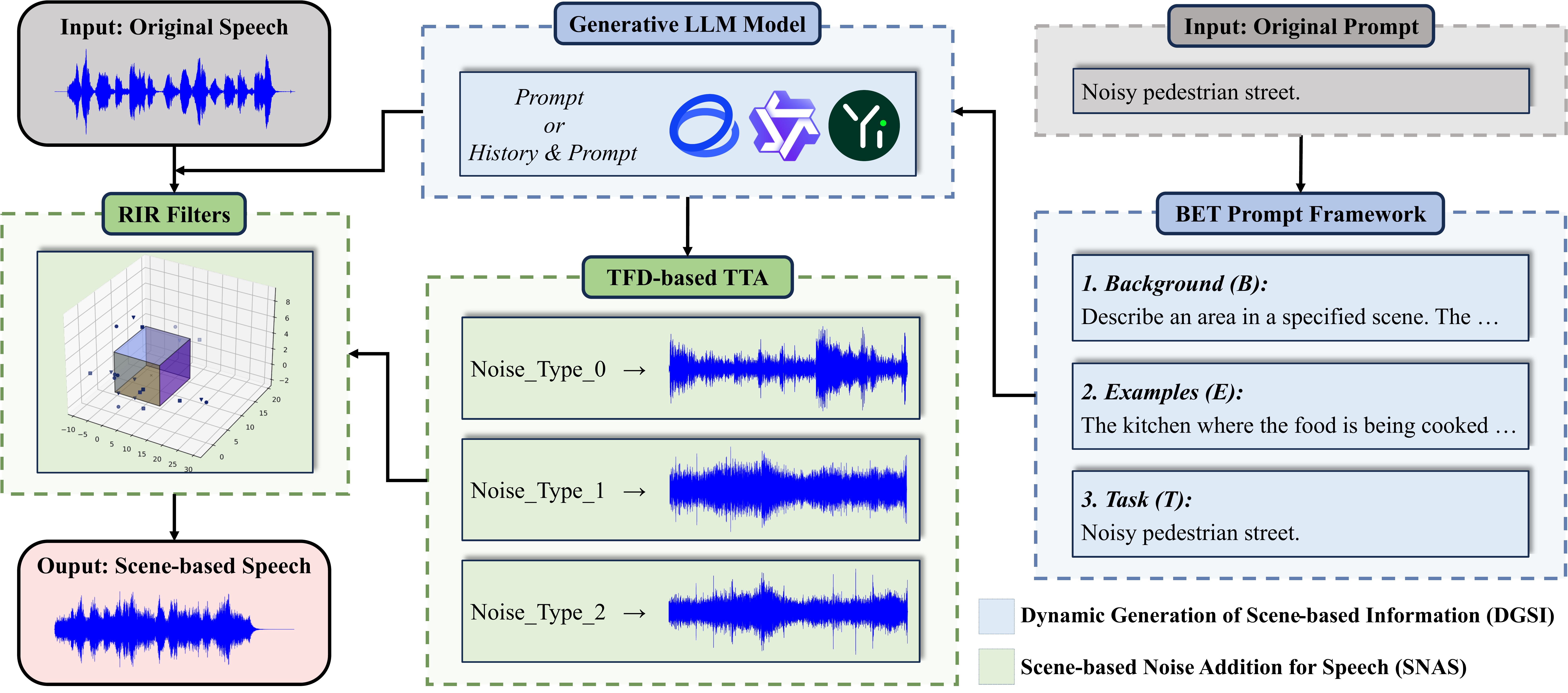}
\end{graphicalabstract}

\begin{highlights}
    \item By leveraging a textual description of the target environment within a BET (Background, Examples, Task) prompt framework, the DGSI module automates scene enumeration and description, enabling highly accurate and dynamic scene simulations.
    \item The SNAS module leverages a Time-Frequency Diffusion-based (TFD) Text-To-Audio (TTA) model to synthesize scene-specific noise, which is then fused with clean speech via RIR filters. This approach automates the traditionally labor-intensive process of replicating diverse and complex acoustic environments. 
    \item We rigorously evaluate DGSNA across three key dimensions: a comparative analysis against baseline noise-addition methods, an assessment of its impact on ASR and KWS performance across varying Added Noise Rates (ANR), and a study of its generalizability across diverse LLMs. These findings are further substantiated by detailed case studies that demonstrate the framework’s reliability.
\end{highlights}


\end{frontmatter}

\section{Introduction}
\label{sec1}

\begin{figure}[ht]
    \centering
    \includegraphics[width=.6\textwidth]{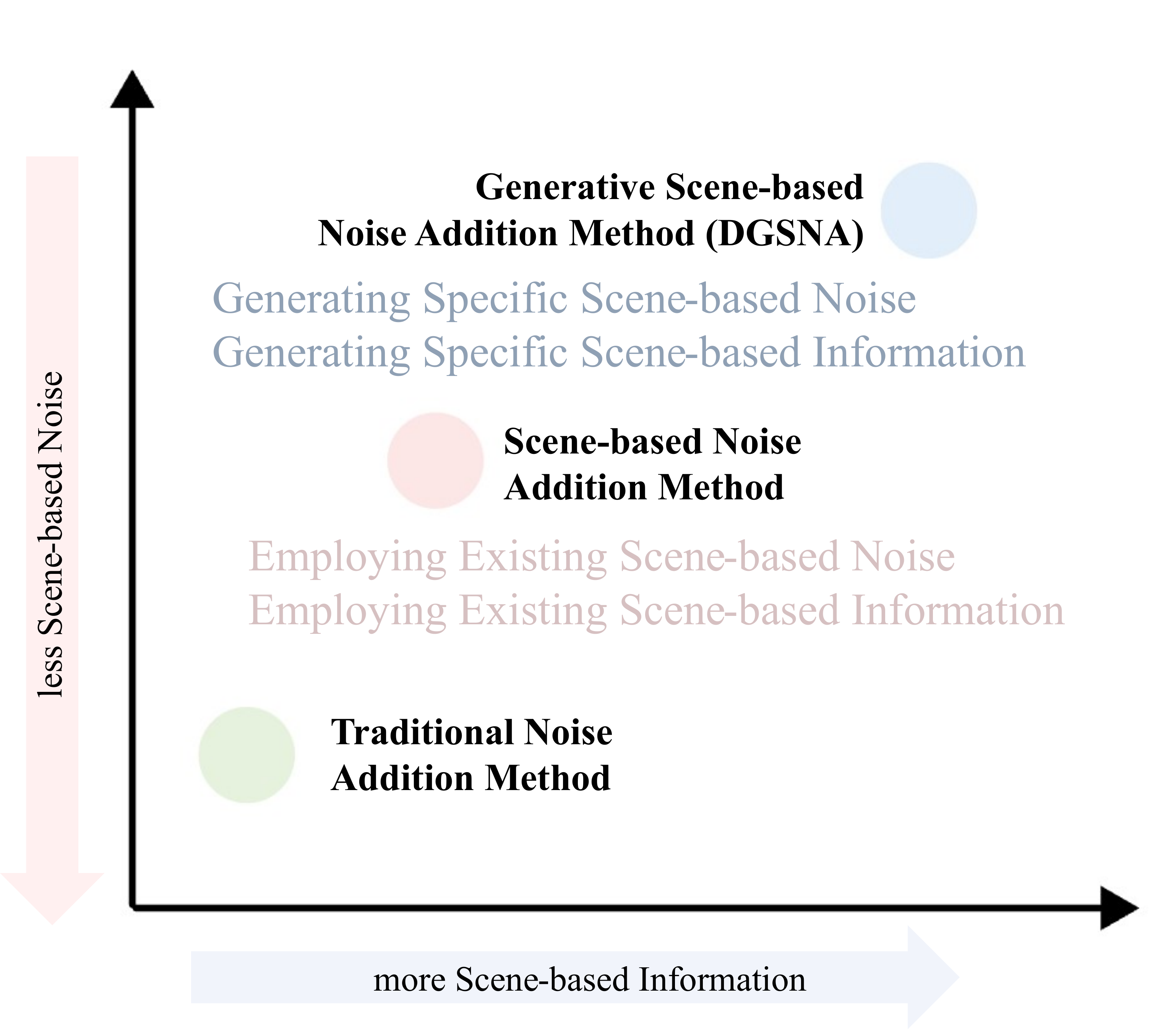}
    \caption{Comparative analysis of different noise addition methods for speech data augmentation.}\label{fig1}
\end{figure}

Synthesizing noisy speech datasets via noise addition is fundamental to improving performance in speech-centric tasks, such as Automatic Speech Recognition (ASR) and Keyword Spotting (KWS). The construction of scene-based noise~\citep{ref1} is essential for replicating authentic acoustic environments, thereby increasing the effectiveness of noise-augmentation strategies. By providing more diverse and realistic simulations, these methods bolster the robustness of downstream models in challenging scenarios—such as those involving the cocktail party effect~\citep{ref2, ref3}.

Figure~\ref{fig1} illustrates a stage-by-stage comparison of various noise addition methods. To accurately reproduce a noisy environment, both scene-based information and noise are essential. The former encompasses scene dimensions, microphone placement, and the types and locations of both noise and voice sources. The latter must align with these specified noise types and can be further diversified across a wide range of acoustic levels.

Traditional noise addition for speech, which includes generating random noise, simulating with filters, and modifying spectral properties, is effective for speech data augmentation but cannot incorporate specific environmental noises, limiting its utility in targeted acoustic environments.
To address this, scene-based noise addition has emerged, generally following two approaches: physically constructing acoustic environments for direct recording or applying Room Impulse Response (RIR) filters to simulate spatial characteristics~\citep{ref1}. Because physical modeling is often time- and resource-prohibitive, RIR simulation has become the prevailing standard~\citep{ref5,ref6}.

Despite these advancements, non-generative scene-based methods remain limited by their reliance on pre-recorded noise and static simulation parameters. While effective in controlled settings, they struggle to generalize to unforeseen environments. Because real-world acoustic diversity cannot be exhaustively cataloged, the labor-intensive process of replicating these scenes creates a data bottleneck, leaving researchers with insufficient variety for robust model training.

To address these challenges, we propose Dynamic Generative Scene-based Noise Addition (DGSNA), which integrates Dynamic Generation of Scene-based Information (DGSI) with Scene-based Noise Addition for Speech (SNAS). 
Our implementation of DGSNA offers the following key contributions:
\begin{itemize} 
    \item By leveraging a textual description of the target environment within a BET (Background, Examples, Task) prompt framework, the DGSI module automates scene enumeration and description, enabling highly accurate and dynamic scene simulations.
    \item The SNAS module leverages a Time-Frequency Diffusion-based (TFD) Text-To-Audio (TTA) model to synthesize scene-specific noise, which is then fused with clean speech via RIR filters. This approach automates the traditionally labor-intensive process of replicating diverse and complex acoustic environments.
    \item We rigorously evaluate DGSNA across three key dimensions: a comparative analysis against baseline noise-addition methods, an assessment of its impact on ASR and KWS performance across varying Added Noise Rates (ANR), and a study of its generalizability across diverse LLMs. These findings are further substantiated by detailed case studies that demonstrate the framework’s reliability.
\end{itemize}

\section{Related work}
\label{sec2}

\subsection{Traditional noise addition methods}
\label{subsec2_1}
Traditional noise addition methods mainly include random noise generation, filter simulation, and spectral transformation. 
Random noise generation creates random values following specific distributions~(e.g., Gaussian~\citep{ref10}, uniform~\citep{ref11}, and those generated by Linear Feedback Shift Register~(LFSR) algorithms~\citep{ref12}), to mimic noise signal amplitude and frequency features. 
Filter simulation uses digital filters to reproduce specific noise types, including white~\citep{ref13}, pink~\citep{ref14}, and brown noise~\citep{ref15}, and also permits customizing filter frequency responses to achieve desired noise characteristics. 
Spectral transformation utilizes Fourier transforms~\citep{ref16} to add specific frequency noise to the original speech signal in the frequency domain. 
This method also converts original speech spectrum elements into noise with minimal distortion, utilizing techniques such as spectral blur~\citep{ref17} and SpecAugment~\citep{ref18}.

However, these conventional methods primarily focus on general speech data augmentation rather than optimizing performance for specific acoustic environments.
To overcome this limitation, researchers have developed scene-based noise addition techniques.

\subsection{Scene-based noise addition methods}
\label{subsec2_2}
Scene-based noise addition methods are broadly categorized into two approaches: physical scene construction and RIR simulation. 
The latter has become increasingly prevalent, as the substantial time and resource demands of real-world recording often make physical data collection impractical.
However, simulating effective RIR filters presents challenges. 
Image Source Method~(ISM)~\citep{ref19}, calculates sound propagation by mirroring virtual \textit{image sources} across room boundaries. ISM typically assumes an empty, parallelepiped or rectangular room with a fixed absorption rate for all boundaries. 
However, this simplified approach often fails to capture the complex acoustics of real-world environments, where reflections are heavily influenced by spatial geometry and material properties. Consequently, numerous sophisticated methods have emerged to enhance the quality and realism of RIR filters.
Ray-tracing strategies have been explored for computing sound paths using explicit room models~\citep{ref21}, while diffuse-based methods approximate late reverberation~\citep{ref22,ref23}. 
Furthermore, neural networks~\citep{ref24}, notably Generative Adversarial Networks~(GANs)~\citep{ref25}, have been employed to refine RIR filter simulation, aiming to more closely approximate the distribution of real recorded RIR filters~\citep{ref26}.

Current research on scene-based noise addition largely focuses on known or predictable scenes, often referred to as non-generative scene-based noise addition. 
A principal limitation of these methods is their inability to capture the extensive diversity of real-world scenes, which exceeds practical cataloging and description capabilities. Moreover, the creation or simulation of known scenes and recording scene-based noise are both labor- and time-intensive processes. This complexity constrains researchers' efforts to collect sufficient scene-based speech data for effective model training. Inspired by the success of generative LLM models across text~\citep{ref27}, audio~\citep{ref28}, and image modalities~\citep{ref29}, this paper introduces DGSNA, a novel generative scene-based noise addition method.

\subsection{Generative Text-to-Audio Models}
\label{subsec2_3}
Once the scene-specific parameters are established, generative TTA models can synthesize background noise that aligns with these specifications. TTA is an emerging field focused on generating diverse audio samples from natural language prompts. Early TTA models~\citep{ref37,ref38} primarily relied on one-hot labels, which restricted the label space and limited generative variety, often resulting in monotonous outputs. In contrast, natural language descriptions provide richer, more nuanced information for audio synthesis.

Recent breakthroughs in diffusion-based generative models~\citep{ref39,ref40} have demonstrated remarkable capabilities in both content comprehension and creation. Diffsound~\citep{ref9}, the first diffusion-based TTA model, significantly outperformed its predecessors by generating quantized discrete tokens from mel-spectrograms. This was followed by AudioGen~\citep{ref41}, which utilized an autoregressive model in a discrete waveform space to surpass Diffsound’s performance. Subsequently, AudioLDM~\citep{ref42} pioneered the use of continuous Latent Diffusion Models (LDMs), achieving superior audio quality and efficiency compared to discrete token-based systems. Other leading models, such as Tango~\citep{ref44} and Make-an-Audio~\citep{ref45}, have since adopted LDMs to refine latent space denoising. Building on these advancements, this paper proposes SNAS, a framework that integrates TTA-generated environmental noise into speech signals to produce realistic, scene-aware speech.

\section{Methodology}
\label{sec3}
\begin{figure}[ht]
    \centering
    \includegraphics[width=.96\textwidth]{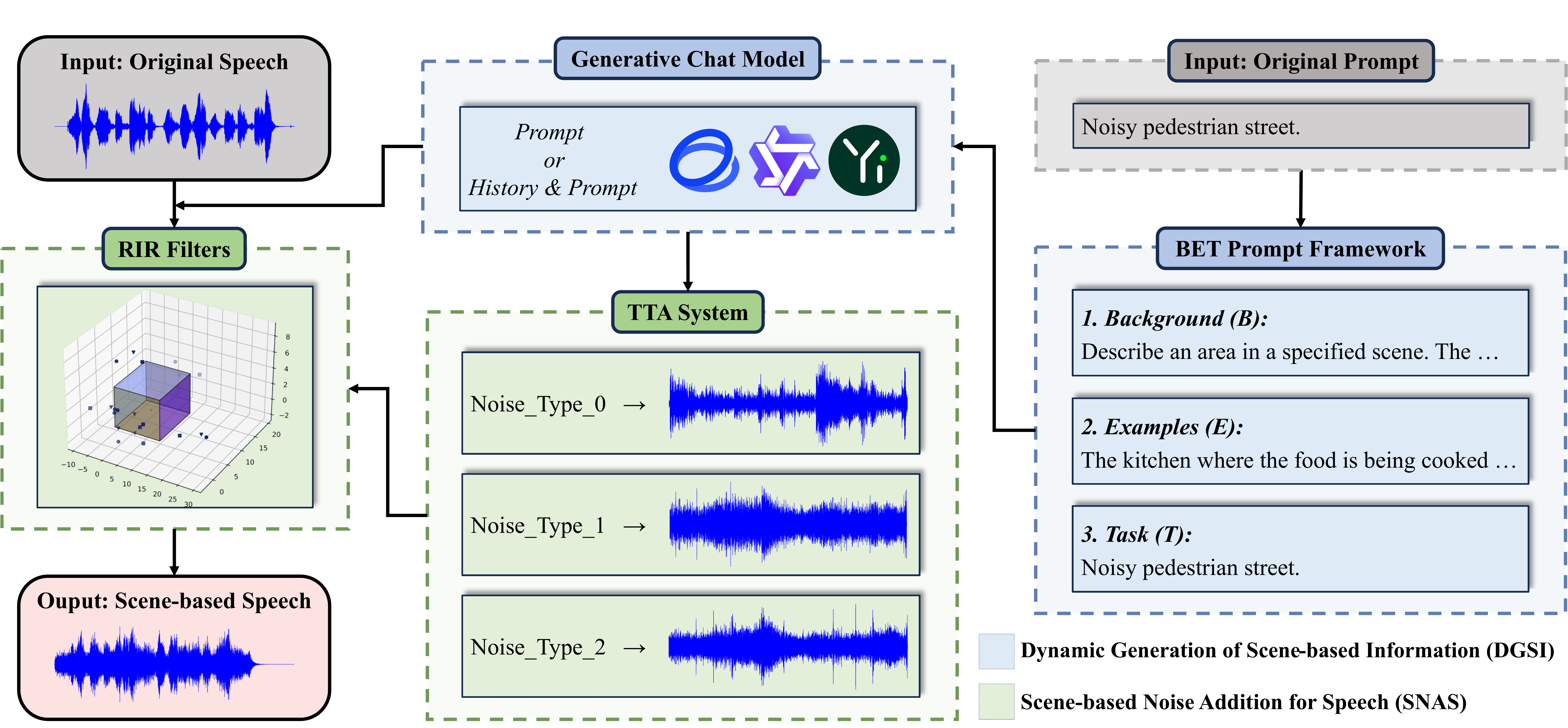}
    \caption{
    The overall framework of the proposed DGSNA method is illustrated below. Within this structure, the BET Prompt Framework for the DGSI module comprises three components: Background (B), Examples (E), and Task (T). Additionally, the TFD-based TTA component utilizes a Time-Frequency Collaborative Audio Encoder to generate scene-specific noise.
    }
    \label{fig2}
\end{figure}

A comprehensive overview of the DGSNA is provided in Figure \ref{fig2}. 
In the DGSI module, the text description of the target scene is integrated into the BET prompt framework. This process generates a BET prompt that is subsequently fed into generative LLM models to dynamically generate scene-based information.
Within the SNAS module, the process is initiated by generating scene-based noise through TFD-based TTA models based on the types of scene-based noise identified in the scene-based information. Subsequently, RIR filters are used to combine this scene-based noise with the scene-based information and the original speech, thus providing comprehensive scene-based speech.

\subsection{DGSI: Dynamic Generation of Scene-based Information}
\label{subsec3_1}

\begin{figure}[!ht]
    \centering
    \includegraphics[width=.86\textwidth]{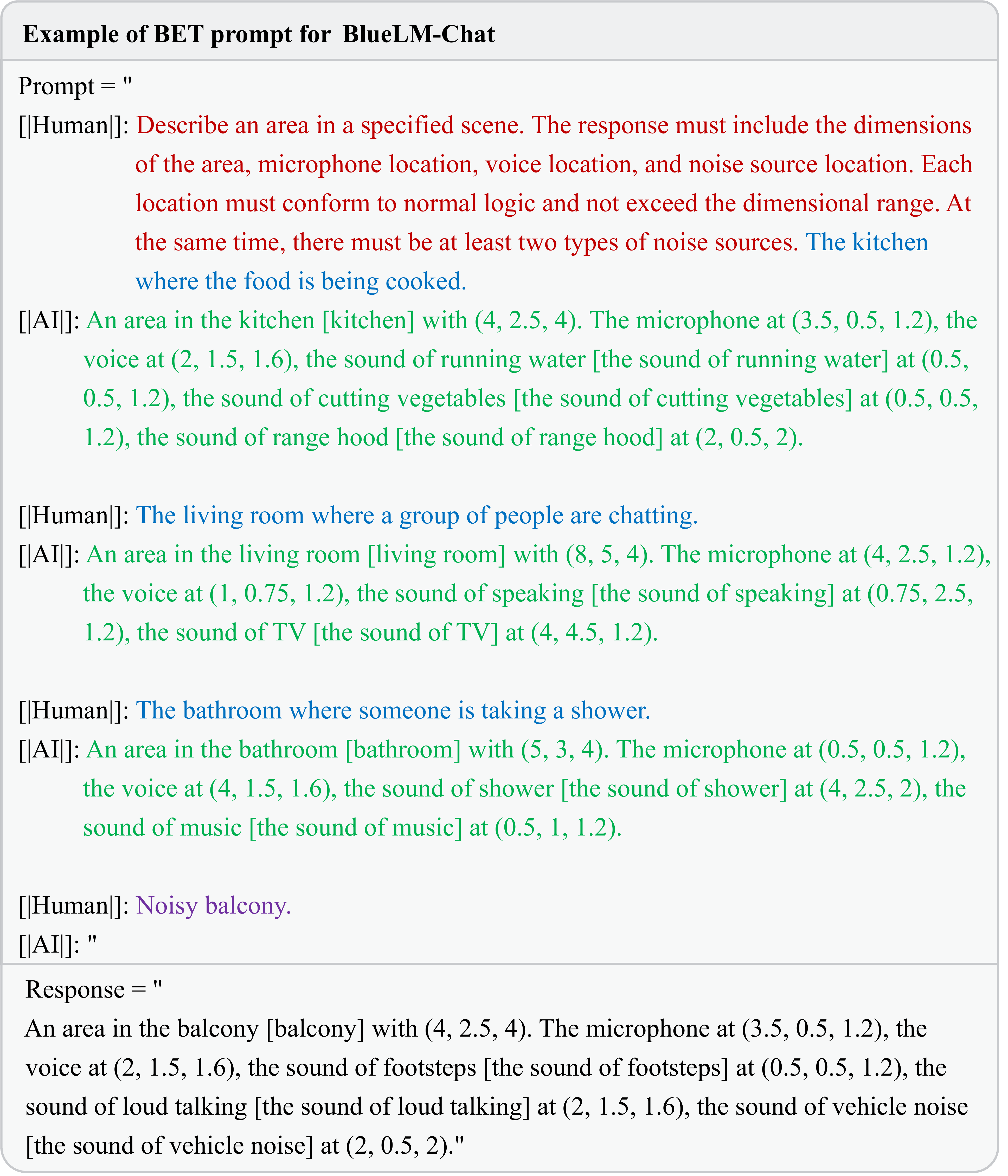}
    \caption{Examples of scene dynamic generation. Initially, the \textit{B}~(Background) component involves the user providing a clear and detailed description of the specified scene's design and context~(red font). Subsequently, the \textit{E}~(Examples) component requires the user to input few-shot prompts, including a text description of the target scene~(blue font). Using these inputs, the generative LLM model generates scene-based information that aligns with the predefined task background~(green font). Finally, the \textit{T}~(Task) component involves the user specifying requirements for the dynamic generation of the scene~(purple font).}
\label{fig3}
\end{figure}

Using the BET prompt framework, the DGSI component leverages generative LLMs to synthesize dynamic, scene-specific information for diverse acoustic environments.
An example of scene dynamic generation using a generative LLM model is provided in Figure \ref{fig3}. 
To facilitate the production of specific information about scenes by the generative LLM model, this paper introduces a new prompt framework, BET, which is an acronym for Background~(\textit{B}), Examples~(\textit{E}), and Task~(\textit{T}).

In summary, the \textit{E} component of the BET utilizes a structured query-and-response format to guide the generative LLM model in producing detailed scene-based information. 
The Query can be represented as $\textbf{Q}_n=[a_n,s_n]$, where $n$ denotes the current example number, $a$ is an adjective describing the scene~(e.g., \textit{Noisy} in Figure \ref{fig3}), and $s$ represents the type of the scene~(e.g., \textit{balcony}). 
The Response can be represented as $\textbf{R}_n=[d_n,s_n,{ml}_n,{sl}_n,{nt}_n^1,{nl}_n^1,{nt}_n^2,{nl}_n^2,...,{nt}_n^m,{nl}_n^m]$, where $m$ denotes the number of noise types included in the response, $d$ refers to the scene dimensions~(e.g., \textit{(4, 2.5, 4)}), $ml$ is the microphone location~(e.g., \textit{(3.5, 0.5, 1.2)}), $sl$ is the speaker location~(e.g., \textit{(2, 1.5, 1.6)}), $nt$ represents the noise type~(e.g., \textit{the sound of footsteps}), and $nl$ is the noise location~(e.g., \textit{(0.5, 0.5, 1.2)}). The T can be represented as $\textbf{T}=[a,s]$. Finally, the output from the generative LLM model, which constitutes the dynamically generated scene-based information, can be represented as scene-based information $\textbf{SI}=[d,s,ml,sl,{nt}^1,{nl}^1,{nt}^2,{nl}^2,...,{nt}^m,{nl}^m]$.

\subsection{SNAS: Scene-based Noise Addition for Speech}
\label{subsec3_2}
\subsubsection{Time Frequency Collaborative Audio Encoder}

\begin{figure}[!ht]
    \centering
    \includegraphics[width=.96\textwidth]{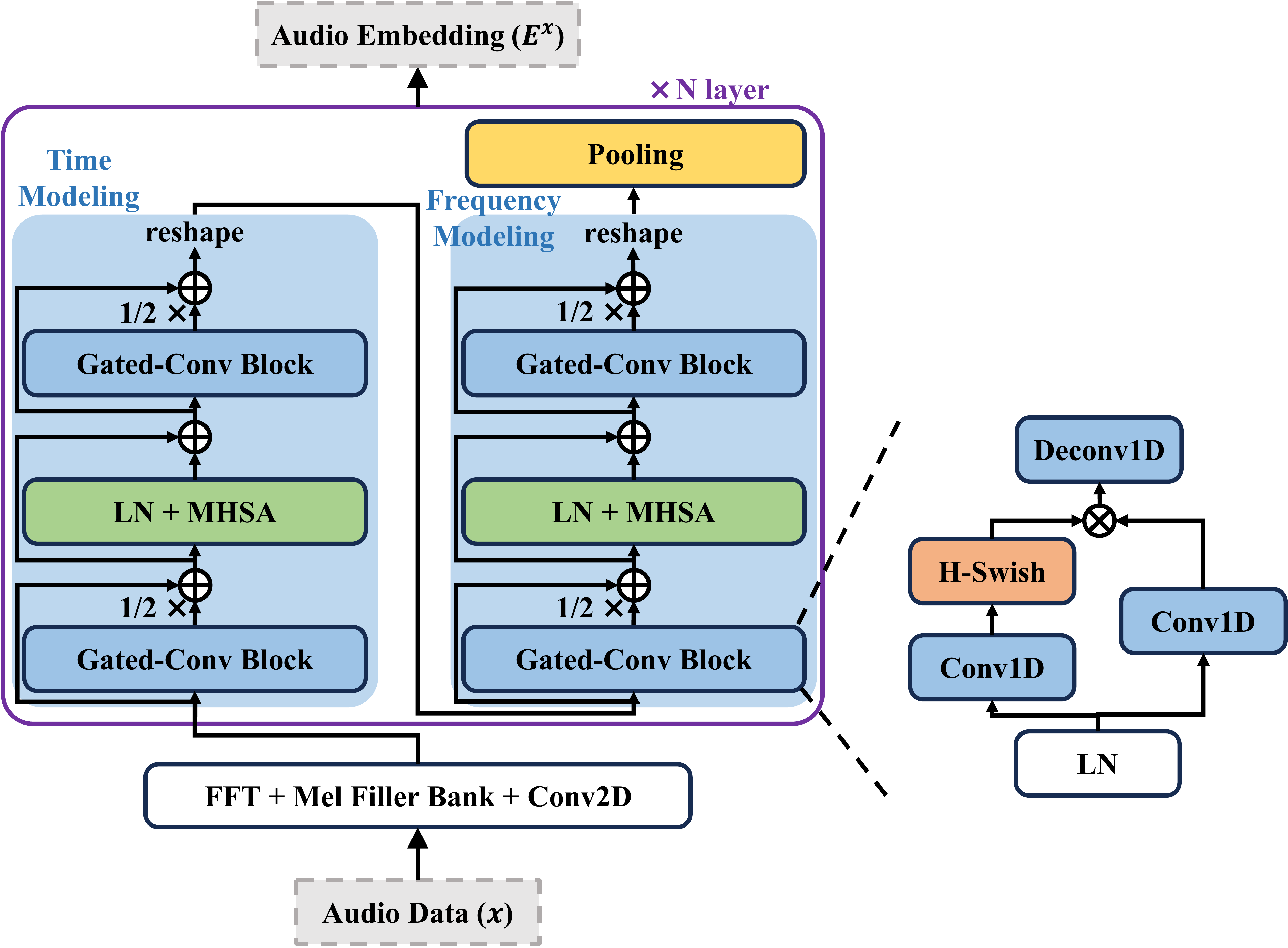}
    \caption{Architecture of the Time-Frequency Collaborative Audio Encoder within the diffusion-based TTA framework.}
    \label{audio_encoder}
\end{figure}

To extract high-order audio representations, this work introduces a Time-Frequency Collaborative (TFC) Audio Encoder (Figure \ref{audio_encoder}). The architecture comprises $N$ identical, cascaded layers, each sequentially performing feature pooling, frequency-domain modeling, and time-domain modeling.

Let $x$ represent the one-dimensional raw audio stream of the input. Using a combination of the Fast Fourier Transform (FFT) and 2D convolution, the model first reduces dimensions and maps features, producing the initial time-frequency features $X_{0} = \text{Conv2D}(\text{FFT}(x))$.

After receiving the output $H_{l-1}$ from the preceding layer (for the first layer, $H_{0} = X_{0}$), each encoder layer progressively goes through the following three fundamental stages.

This step prioritizes the temporal evolution of the audio signal. To facilitate local augmentation, features are processed through a gated convolutional block.
Next, they go through a multi-head self-attention mechanism with residual connections to capture long-range dependencies. Lastly, they go through another gated convolutional block to deepen the feature representation. To prepare it for frequency-domain processing, its output $H_{time}$ is reshaped to change the dimension:

\begin{equation}
    \hat{z}_{t} = H_{l-1} + \frac{\text{GatedConv}(H_{l-1})}{2}
\end{equation}
\begin{equation}
    \tilde{z}_{t} = \text{LN}(\hat{z}_{t} + \text{MHSA}(\hat{z}_{t}))
\end{equation}
\begin{equation}
    H_{time} = \text{Reshape}(\tilde{z}_{t} + \frac{\text{GatedConv}(\tilde{z}_{t})}{2})
\end{equation}

The time-domain and frequency-domain modeling units exhibit structural symmetry. Specifically, the frequency-domain unit iteratively applies a sequence of operators across the frequency axis upon receiving the reshaped time-domain features, $H_{time}$. The model can alternately optimize temporal and frequency resolution thanks to this design:

\begin{equation}
    \hat{z}_{f} = H_{time} + \frac{\text{GatedConv}(H_{time})}{2}
\end{equation}
\begin{equation}
    \tilde{z}_{f} = \text{LN}(\hat{z}_{f} + \text{MHSA}(\hat{z}_{f}))
\end{equation}
\begin{equation}
    H_{freq} = \text{Reshape}(\tilde{z}_{f}) + \frac{\text{GatedConv}(\tilde{z}_{f})}{2}
\end{equation}

The final representation of each modeling layer, $H_{l} = \text{Pooling}(H_{freq})$, is produced by the model using a pooling layer at the end to selectively compress and reduce the dimensionality of the complex features retrieved from both domains.

The gated convolutional blocks described above employ a dual-branch architecture, as illustrated in the detail diagram on the right.
Following layer normalization, element-wise multiplication is used to filter the input features $z$:
\begin{equation}
    \text{GatedConv}(z) = \text{Deconv}(\text{H-Swish}(\text{Conv1D}_1(\text{LN}(z))) \otimes \text{Conv1D}_2(\text{LN}(z)))
\end{equation}

The final output features are transformed into a fixed-dimensional audio embedding vector $E^x = H_{N}$ for downstream tasks, following $N$ layers of deep modeling.

\subsubsection{Scene-based noise generation}

\begin{figure}[!ht]
    \centering
    \includegraphics[width=.9\textwidth]{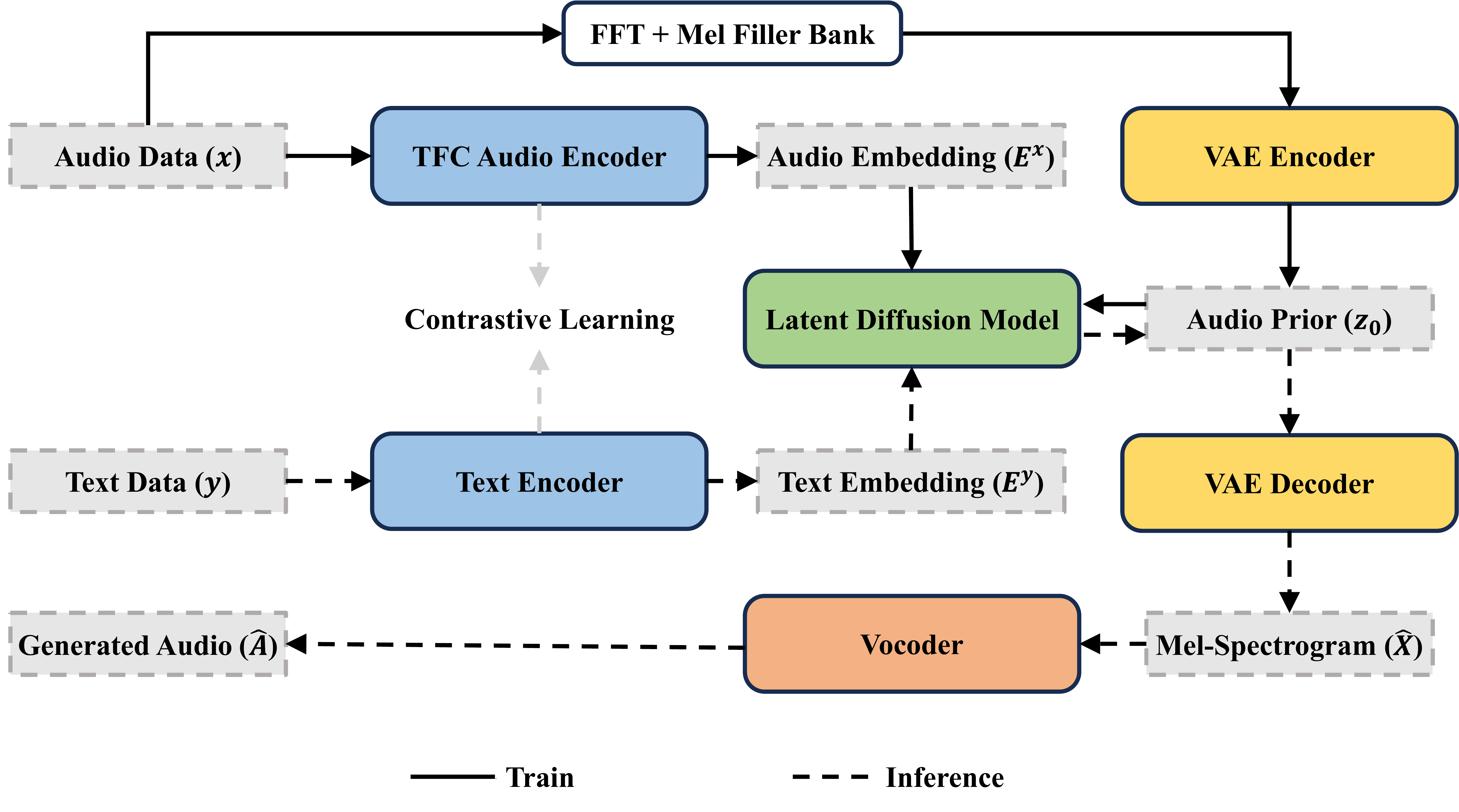}
    \caption{Architecture of the proposed TFD-based TTA framework.}
    \label{TTA_overall_framework}
\end{figure}

A TFD-based TTA model with TFC Audio Encoder is proposed in this study (Figure \ref{TTA_overall_framework}). From the \textbf{SI} produced by DGSI, SNAS extracts noise types $\textbf{NT}=[{nt}^1,{nt}^2,...,{nt}^m]$ as prompt text to instruct the TTA model to produce context-aware noise. The specific procedure is as follows: each noise type ${nt}^i$ is utilized as input to the TTA model once an NT list has been extracted. A previous representation of the audio $\textbf{E}^y$ is then produced via LDM utilizing the Contrastive Language-Audio Pretraining (CLAP) \citep{ref46} technique. A Variational Auto-Encoder (VAE) \citep{ref63} decoder reconstructs this previous representation into a Mel-spectrum, which is then transformed into the final audio waveform using the Hifi-GAN \citep{ref48} encoder. 
By generating noise that closely aligns with the target scenario, this integrated approach significantly enhances the realism and immersive quality of the contextualized speech.

The CLAP model is used as the text encoder $f_{\textrm{text}}(\cdot)$ to translate text descriptions $y$ to a cross-modal latent space $\textbf{E}^y$. Using the text embedding $\textbf{E}^y$ as a guide, the LDM model gradually reconstructs the prior representation of the audio $z_0$ using a reverse denoising process throughout the inference step, beginning with a Gaussian distribution $p(z_N) \sim \mathcal{N}(0, I)$. The following is the definition of the probability distribution $p_\theta$:

\begin{equation}
    p_\theta(z_{n-1}|z_n, E^y) = \mathcal{N}\Big(z_{n-1}; \mu_\theta(z_n, n, E^y), \sigma_n^2 I\Big)
\end{equation}

The mean and variance are parameterized as follows:

\begin{equation}
    \mu_\theta(z_n, n, E^y) = \frac{1}{\sqrt{\alpha_n}} \left( z_n - \frac{\beta_n}{\sqrt{1-\overline{\alpha}_n}} \epsilon_\theta(z_n, n, E^y) \right)
\end{equation}

\begin{equation}
    \sigma_n^2 = \frac{1-\overline{\alpha}_{n-1}}{1-\overline{\alpha}_n} \beta_n
\end{equation}

The predicted noise in the denoising process is denoted by $\epsilon_\theta(z_n,n,\textbf{E}^y)$, $\alpha_n = 1-\beta_n$, and $\overline{\alpha}_n$ is defined as $\overline{\alpha}_n=\prod_{i=1}^{n}\alpha_i$. A predetermined set of hyperparameters is $\beta_n$.

\subsubsection{Scene-based noise addition}
Following the methodology outlined in PyRoomacoustics \citep{ref21}, we
model the environment as a shoebox-shaped enclosure based on the scene information.
First, the \textbf{SI} produced by DGSI is used to extract parameters such as scene dimensions $d$, microphone location $ml$, speaker location $sl$, and noise location $nl$. The generated scene-based noise $\textbf{SN}=[\hat{A}^1,\hat{A}^2,...,\hat{A}^m]$ is then added to these parameters and used as input for the RIR filter. The microphone is situated at $l = ml$, the simulated room dimensions are $d$, and the actual sound source $s_0$ can be either the original speech (placed at $sl$) or a noise source, i.e., $s_0=[\hat{A}^i,{nl}^i]$.

The procedure begins by mapping the source audio to spatial coordinates within a virtual 3D environment, reconstructed from the extracted feature set.
The ISM is then employed to discretize the resulting sound field.
A set of visible mirror sources $V_l~(s_0)$ is identified by computing the locations of virtual pictures created by sound wave reflections off the walls. The RIR between the sound source $s_0$ and the microphone at location $l$ can be written as follows:

\begin{equation}
    a_r(s_0, n) = \sum_{s \in V_l(s_0)} \frac{\gamma_s (1-\alpha)^{R(s)}}{4\pi \|l-s\|} \delta_{LP} \left( n - F_s \frac{\|l-s\|}{c} \right)
\end{equation}

where $c$ is the speed of sound in air, $\delta_{LP}$ is the windowed low-pass Sinc function, $n$ is the current time step, $F_s$ is the sampling rate, $R(s)$ is the reflection order of the sound source $s$, $\gamma_s \in [0,1]$ is the amplitude of the sound source $s$, and $\alpha \in [0,1]$ is the sound absorption coefficient of the wall.

\begin{equation}
    \delta_{LP}(t) =
    \begin{cases}
        \frac{1}{2} \left( 1 + \cos\left( \frac{2\pi t}{T_w} \right) \right), & -\frac{T_w}{2} \le t \le \frac{T_w}{2} \\
        0, & \textrm{otherwise}
    \end{cases}
\end{equation}

Specifically, the window width $T_w$ is a crucial parameter for balancing signal reconstruction accuracy against computational overhead. The final context-aware speech is obtained by convolving the target speech with the calculated RIR.

\section{Experimental settings}
\label{sec4}

This section is organized into five components: benchmark datasets, baseline models, generative LLMs, evaluation metrics, and experimental parameters.

\subsection{Benchmark datasets}
\label{subsec4_1}
To evaluate DGSNA's effectiveness in ASR and KWS tasks, we trained our models on clean datasets to simulate laboratory conditions using noisy datasets to mimic real-world scenarios.

For the ASR experiment, we used Aishell-1~\citep{ref49} as the training set and WenetSpeech~\citep{ref50} as the test set.

For the KWS domain, we were unable to locate datasets analogous to Aishell-1 and WenetSpeech~(i.e., a clean training set and a noisy test set). Therefore, we used the HeySnips~\citep{ref51} dataset, which contains background noise in both training and testing sets, for our KWS experiments.

\begin{itemize}
\item \textbf{Aishell-1:} Aishell-1, released by Beijing Shell Shell Technology Co., Ltd., comprises over 170 hours of Mandarin speech data from 400 speakers. It is a subset of the 500-hour multi-channel AISHELL-ASR0009 corpus, designed for various speech and speaker processing tasks.
\item \textbf{WenetSpeech:} WenetSpeech is a multi-domain Mandarin corpus comprising over 22,400 hours of speech, including 10,000+ hours of high-quality labeled speech, 2,400+ hours of weakly labeled speech, and approximately 10,000 hours of unlabeled speech. 
WenetSpeech comprises three subsets: (a) Dev, which is tailored for speech tools requiring cross-validation; (b) Test\_Net, a matched test set collected from the internet; and (c) Test\_Meeting, a mismatched test set characterized by far-field, conversational, and spontaneous meeting recordings.
\item \textbf{HeySnips:} The keyword "Hey Snips"~(pronounced without a pause) was used. The dataset includes diverse English accents and recording environments, comprising approximately 11,000 wake-word utterances and 86,500~(96 hours) negative examples.
\end{itemize}

During data preprocessing, DGSNA was applied probabilistically; in other words, noise augmentation was not performed on every speech sample. The probability of adding noise in this process is referred to as the Add Noise Rate~(ANR).

\subsection{Baseline models}
\label{subsec4_2}

We employed the Conformer \citep{ref52} as our baseline for ASR experiments via the WeNet \citep{ref53} framework. For KWS, we utilized the Multi-scale Depthwise Temporal Convolution (MDTC) \citep{ref54} model within the WeKws \citep{ref55} framework.

\begin{itemize}
\item \textbf{Conformer:} Conformer adds a Convolution module based on transformer~\citep{ref56} so it can capture both local and global context and get better results on different ASR tasks.
\item \textbf{WeNet:} WeNet implements U2, a novel two-pass approach that unifies streaming and non-streaming end-to-end~(E2E) speech recognition within a single model. 
The U2 architecture comprises three core components: a shared Conformer encoder, a CTC decoder, and a Transformer-based attention decoder.
\item \textbf{MDTC:} MDTC explicitly fuses multi-scale features from different hidden layers with different receptive fields and models long-range temporal features with efficient dilated depthwise temporal convolution.
\item \textbf{WeKws:} WeKws consists of four parts, starting with a global Cepstral Mean and Variance Normalization~(CMVN) layer to normalize the input acoustic features to a normal distribution. 
This is followed by a linear layer that projects the input features into the target dimensionality.
The model architecture consists of a backbone network followed by several binary classifiers. Specifically, we implemented MDTC as our primary backbone.
\end{itemize}

\subsection{generative LLM models}
\label{subsec4_3}

This paper references C-Eval~\citep{ref57}, an evaluation benchmark specifically designed to assess the capabilities of generative LLM models in both Chinese and English contexts.
To accommodate the computational constraints common in research environments, models were strategically selected based on their performance in the C-Eval rankings.
Consequently, the experiment included the following generative LLM models: BlueLM-7B-Chat~\citep{ref58}, Yi-6B-Chat~\citep{ref59}, ChatGLM3-6B~\citep{ref60}, and Qwen-7B-Chat~\citep{ref32}.

\subsection{Evaluation metrics}
\label{subsec4_4}
When evaluating the performance of our speech processing systems, we utilize distinct metrics tailored to the objectives of each individual task. 

For ASR experiments, we employ the Word Error Rate~(WER), which provides a comprehensive measure of transcription accuracy by quantifying the number of substitutions, insertions, and deletions required to align the system's output with the reference transcript. A lower WER signifies better ASR performance. 

In contrast, our KWS experiments focus on detection accuracy, evaluated via the False Rejection Rate (FRR) and False Alarms per Hour (FA/Hour). The FRR indicates the percentage of times the system fails to detect a spoken keyword when it is present, while the FA/Hour measures how frequently the system incorrectly identifies a keyword in the absence of one. These KWS metrics provide a clear understanding of the system's ability to reliably detect target keywords while minimizing erroneous detections.

\subsection{Experimental parameters}
\label{subsec4_5}

\begin{figure}[ht]
    \centering
    \includegraphics[width=.96\textwidth]{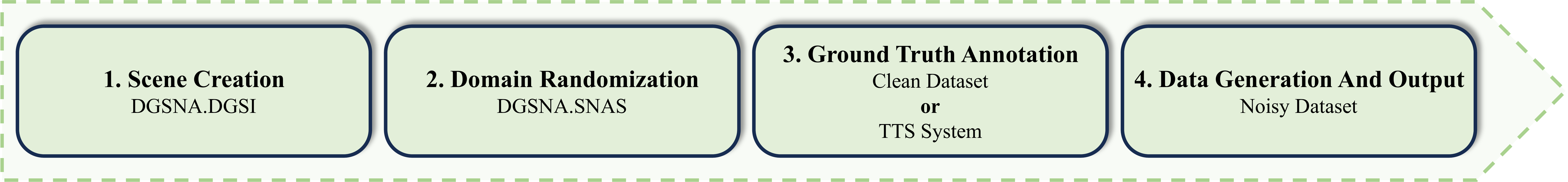}
    \caption{An overview of the DGSNA data generation process and workflow.}\label{fig4}
\end{figure}

The number of diffusion steps~(\textit{ddim\_steps}) for the Denoising Diffusion Implicit Model~(DDIM)~\citep{ref61} were set to 200, controlling the iterations of the denoising process. A guidance scale~(\textit{guidance\_scale}) of 2.5 was employed to steer the model's output based on provided guidance, enhancing alignment with target audio characteristics. The generated audio duration~(\textit{duration}) was set to 5.0 seconds.

In our experiment, PyRoomacoustics was selected as the library to implement the RIR filter. The reverberation time~(\textit{rt60}) is set to 0.5 seconds, representing the time required for the sound pressure level to decrease by 60 decibels~(\textit{dB}). The sampling rate~(\textit{fs}) is configured to 16,000 Hz. The maximum specified order~(\textit{max\_order}) is limited to 1. The reasons for these choices are as follows:

In indoor scenes, where the physical space is constrained, and sound sources are fewer, this setting allows for the simulation of first-order echoes reflecting from the nearest surfaces without overly complicating the acoustic model. This method effectively models essential reflections while preventing the unrealistic accumulation of reverberation that might result from higher-order reflections. Conversely, in outdoor scenes, echoes and reverberations are typically negligible due to the absence of confining structures. In these scenarios, reflections, due to their minimal impact, can conceptually be treated as additional sound sources, simplifying the acoustic modeling process and focusing on direct sound transmission. This analytical approach offers the advantage of simulating extensive acoustic environments without requiring intricate modeling of complex interactions between sound waves and environmental features, particularly beneficial in large, open spaces where such interactions are less defined.

\section{Experimental results}
\label{sec5}

\subsection{DGSNA vs. other noise addition methods}
\label{subsec5_1}

\begin{table}[ht]
  \renewcommand{\arraystretch}{1.8}
  \setlength{\tabcolsep}{3pt}
  \centering
  \scalebox{0.8}{
  \begin{tabular}{l c c c c c c}
    \hline
      \multirow{3}{*}{\textbf{Model}} & \multirow{3}{*}{\textbf{\makecell{Train\\Set}}} & \multirow{3}{*}{\textbf{\makecell{Stream\\Decode}}} & \multicolumn{4}{c}{\textbf{Test WER($\%$) $\downarrow$}} \\
      \cline{4-7}
       & & & \textbf{Aishell-1} & \multicolumn{3}{c}{\textbf{WenetSpeech}} \\
      \cline{4-7}
       & & & \textbf{Test} & \textbf{Dev} & \textbf{Test\_Net} & \textbf{Test\_Meeting} \\
      \hline
      \multirow{2}{*}{\makecell{gaussian\\noise}} & \multirow{2}{*}{Aishell-1} & True & 6.22 & 33.07 & 51.31 & 54.16 \\
       & & False & 5.81 & 32.51 & 50.75 & 53.85 \\
      \hline
      \multirow{2}{*}{SpecAug} & \multirow{2}{*}{Aishell-1} & True & 5.33 & 29.90 & 48.94 & 53.99 \\
       & & False & \textbf{4.93} & \textbf{29.16} & 48.29 & 53.60 \\
      \hline
      \multirow{2}{*}{\makecell{PyRoomacoustics}} & \multirow{2}{*}{Aishell-1} & True & 5.74 & 32.08 & 51.98 & 54.37 \\
       & & False & 5.36 & 31.32 & 51.36 & 54.08 \\
      \hline
      \multirow{2}{*}{DGSNA} & \multirow{2}{*}{Aishell-1} & True & 5.97 & 32.48 & 50.44 & 50.41 \\
       & & False & 5.35 & 31.89 & 49.77 & 50.06 \\
      \hline
      \multirow{2}{*}{\makecell{DGSNA \\ $\&$ \\ SpecAug}} & \multirow{2}{*}{Aishell-1} & True & 5.52 & 29.84 & 46.98 & 48.94 \\
       & & \cellcolor{LightCyan1}False & \cellcolor{LightCyan1}4.96 & \cellcolor{LightCyan1}29.18 & \cellcolor{LightCyan1}\textbf{46.21} & \cellcolor{LightCyan1}\textbf{48.42} \\
      \hline
  \end{tabular}
  }
  \caption{Results of the ASR experiment. All models share the same architecture.}
  \label{tab1}
\end{table}

\begin{figure}[ht]
    \centering
    \includegraphics[width=.8\textwidth]{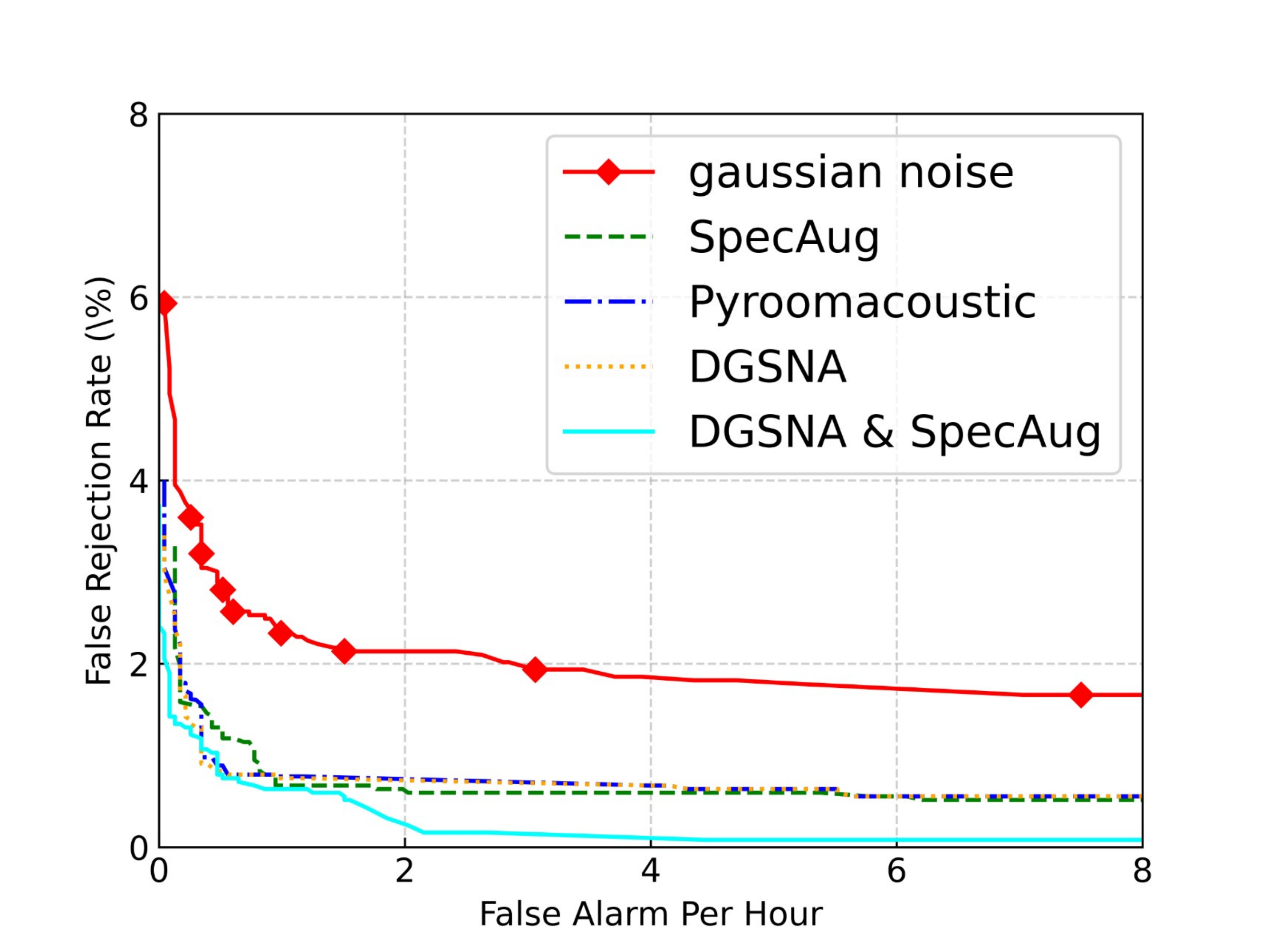}
    \caption{Results of the KWS baseline experiment.}\label{fig5}
\end{figure}

The process of data generation using DGSNA, simplified in Figure \ref{fig4}, highlights that its key difference from other noise addition methods lies in the degree of domain randomization during Step 2. 
To further compare DGSNA with these methods, we conducted comparative experiments in both the ASR and KWS domains, respectively, and the results of the experiments are shown in Table \ref{tab1} and Figure \ref{fig5}, respectively.

The experimental results suggest that DGSNA's effects are similar to those of SpecAugment in practical applications. However, we observed that combining DGSNA and SpecAugment yields even better results. We believe DGSNA adds acoustic environments to clean speech, enabling the model to learn robustness against scene-based noise. Simultaneously, SpecAugment masks portions of the spectrum, improving the model's feature extraction capabilities. These methods are complementary, and to improve model generalizability, we recommend using multiple noise addition methods in combination.

\subsection{Comparison of different TTA models}
\label{subsec5_2}

\begin{table}[!ht]
    \renewcommand{\arraystretch}{1.8}
    \centering
    \begin{tabular}{c c c c c c}
        \hline
        \multirow{3}{*}{TTA} & \multirow{3}{*}{Model} & \multicolumn{4}{c}{Test WER(\%) $\downarrow$} \\
        \cline{3-6}
        & & Aishell-1 & \multicolumn{3}{c}{WenetSpeech} \\
        \cline{3-6}
        & & Test & Dev & Test\_Net & Test\_Meeting \\
        \hline
        Diffsound & DGSNA \& SpecAug & 5.07 & 29.38 & 46.67 & 49.03 \\
        AudioGen & DGSNA \& SpecAug & 4.98 & 29.19 & 46.85 & 49.24 \\
        AudioLDM & DGSNA \& SpecAug & 5.00 & 29.20 & 46.43 & 48.61 \\
        TFD-based TTA(N=2) & DGSNA \& SpecAug & 5.02 & 29.25 & 46.50 & 48.75 \\
        TFD-based TTA(N=4) & DGSNA \& SpecAug & 4.97 & 29.21 & 46.38 & 48.55 \\
        TFD-based TTA(N=6) & DGSNA \& SpecAug & \textbf{4.96} & \textbf{29.18} & \textbf{46.21} & \textbf{48.42} \\
        \hline
    \end{tabular}
    \caption{Results of the ASR ablation experiment}
    \label{ASR_2}
\end{table}

\begin{figure}[!ht]
    \centering
    \includegraphics[width=.8\textwidth]{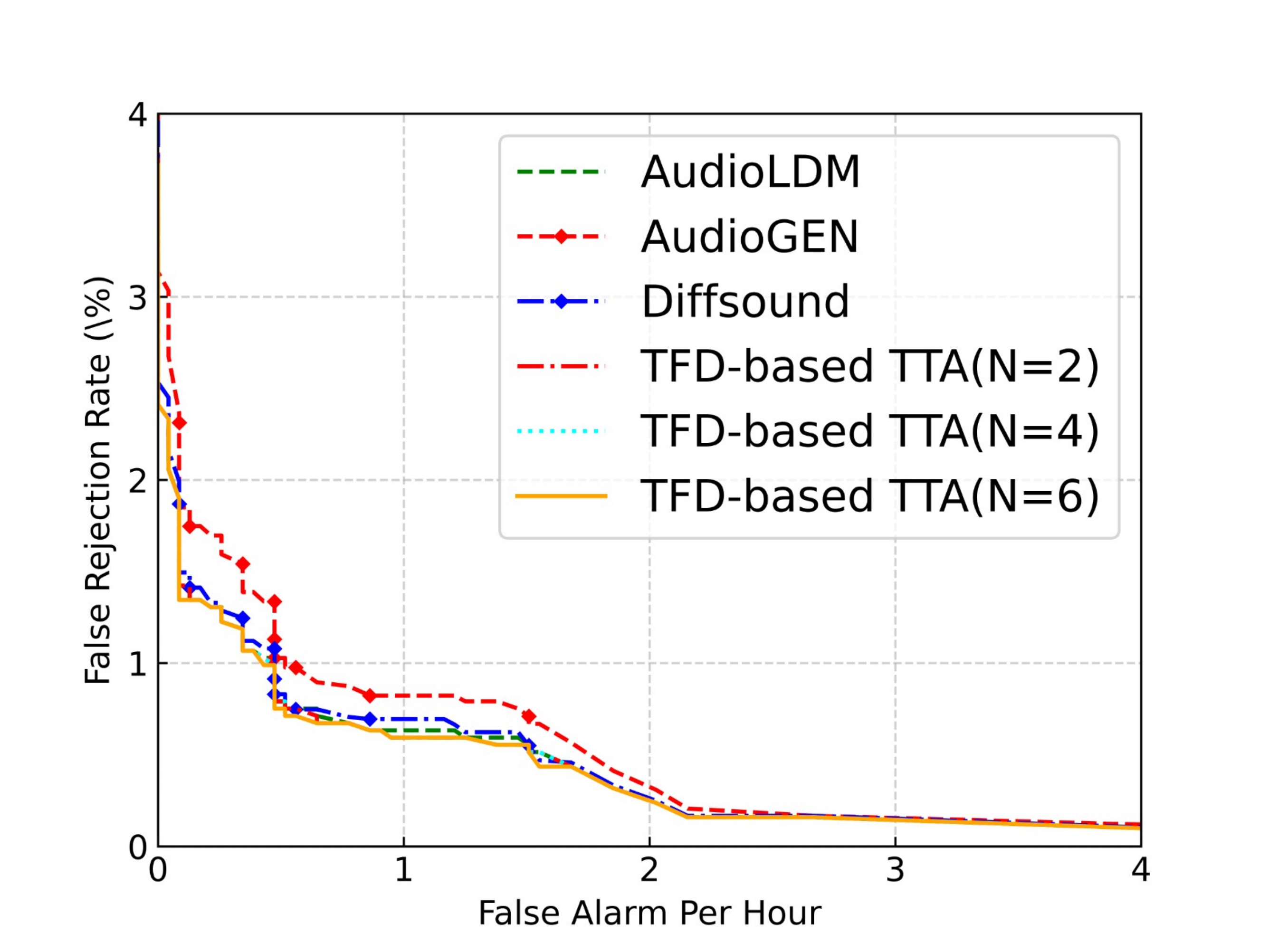}
    \caption{Results of the KWS ablation experiment}
    \label{KWS_2}
\end{figure}

We compared three popular systems: Diffsound, AudioGen, and AudioLDM, in order to determine which audio generation engine would be best for scenario-based noise addition jobs. The TTA model suggested in this work performed the best in both ASR and KWS tasks, according to the experimental data (see Table \ref{ASR_2} and Figure \ref{KWS_2}).

In comparison to systems based on autoregressive or basic GAN structures, the TTA model suggested in this work, which is based on a time-frequency cooperative audio encoder and the LDM architecture, shows more generative diversity and randomness. It can produce environmental noise with better details and more layered textures thanks to this feature. This richness greatly expands the model's acoustic bounds and improves its upper limit for managing unknown disturbances during training by successfully simulating a wider range of "long-tail" noise events.

\subsection{Comparison of different ANR}
\label{subsec5_3}

\begin{table}[!ht]
  \renewcommand{\arraystretch}{1.6}
  \setlength{\tabcolsep}{3pt}
  \centering
  \scalebox{0.75}{
  \begin{tabular}{l c c c c c c c}
    \hline
    \multirow{3}{*}{\textbf{Model}} & \multirow{3}{*}{\textbf{\makecell{Train\\Set\\ANR}}} & \multirow{3}{*}{\textbf{\makecell{Stream\\Decode}}} & \multicolumn{5}{c}{\textbf{Test WER($\%$) $\downarrow$}} \\
    \cline{4-8}
     & & & \multicolumn{5}{c}{\textbf{Aishell-1 ANR}} \\
    \cline{4-8}
     & & & \textbf{0$\%$} & \textbf{10$\%$} & \textbf{20$\%$} & \textbf{30$\%$} & \textbf{40$\%$} \\
    \hline
    \multirow{2}{*}{Baseline} & \multirow{2}{*}{0\%} & True & 5.33 & 9.90 & 14.30 & 19.04 & 23.72 \\
     & & False & \textbf{4.93} & 9.47 & 13.86 & 18.51 & 23.23 \\
    \hline
    \multirow{2}{*}{DGSNA\_10} & \multirow{2}{*}{10\%} & True & ${5.48}_{\color{Firebrick4}{+2.81\%}}$ & ${6.80}_{\color{Cyan4}{-31.31\%}}$ & ${8.47}_{\color{Cyan4}{-40.77\%}}$ & ${9.44}_{\color{Cyan4}{-50.42\%}}$ & ${11.01}_{\color{Cyan4}{-53.58\%}}$ \\
     & & False & ${4.94}_{\color{Firebrick4}{+0.20\%}}$ & ${6.25}_{\color{Cyan4}{-34.00\%}}$ & ${7.83}_{\color{Cyan4}{-43.51\%}}$ & ${8.74}_{\color{Cyan4}{-52.78\%}}$ & ${10.31}_{\color{Cyan4}{-55.62\%}}$ \\
    \hline
    \multirow{2}{*}{DGSNA\_20} & \multirow{2}{*}{20\%} & True & ${5.52}_{\color{Firebrick4}{+3.56\%}}$ & ${6.72}_{\color{Cyan4}{-32.12\%}}$ & ${7.86}_{\color{Cyan4}{-45.03\%}}$ & ${8.99}_{\color{Cyan4}{-52.78\%}}$ & ${10.00}_{\color{Cyan4}{-57.84\%}}$ \\
     & & \cellcolor{LightCyan1}False & \cellcolor{LightCyan1}${4.96}_{\color{Firebrick4}{+0.61\%}}$ & \cellcolor{LightCyan1}$\mathbf{{6.05}_{\color{Cyan4}{-36.11\%}}}$ & \cellcolor{LightCyan1}${7.19}_{\color{Cyan4}{-48.12\%}}$ & \cellcolor{LightCyan1}${8.23}_{\color{Cyan4}{-55.54\%}}$ & \cellcolor{LightCyan1}${9.24}_{\color{Cyan4}{-60.22\%}}$ \\
    \hline
    \multirow{2}{*}{DGSNA\_30} & \multirow{2}{*}{30\%} & True & ${5.69}_{\color{Firebrick4}{+6.75\%}}$ & ${6.73}_{\color{Cyan4}{-32.02\%}}$ & ${7.65}_{\color{Cyan4}{-46.50\%}}$ & ${8.75}_{\color{Cyan4}{-54.04\%}}$ & ${9.60}_{\color{Cyan4}{-59.53\%}}$ \\
     & & False & ${5.14}_{\color{Firebrick4}{+4.26\%}}$ & $\mathbf{{6.08}_{\color{Cyan4}{-35.80\%}}}$ & $\mathbf{{6.89}_{\color{Cyan4}{-50.29\%}}}$ & $\mathbf{{7.98}_{\color{Cyan4}{-56.89\%}}}$ & $\mathbf{{8.86}_{\color{Cyan4}{-61.86\%}}}$ \\
    \hline
    \multirow{2}{*}{DGSNA\_40} & \multirow{2}{*}{40\%} & True & ${6.01}_{\color{Firebrick4}{+12.76\%}}$ & ${6.92}_{\color{Cyan4}{-30.10\%}}$ & ${8.08}_{\color{Cyan4}{-43.50\%}}$ & ${9.19}_{\color{Cyan4}{-51.73\%}}$ & ${10.10}_{\color{Cyan4}{-57.42\%}}$ \\
     & & False & ${5.39}_{\color{Firebrick4}{+9.33\%}}$ & ${6.24}_{\color{Cyan4}{-34.11\%}}$ & ${7.29}_{\color{Cyan4}{-47.40\%}}$ & ${8.36}_{\color{Cyan4}{-54.84\%}}$ & ${9.26}_{\color{Cyan4}{-60.14\%}}$ \\
    \hline
  \end{tabular}
  }
  \caption{Results of the ASR experiment (Aishell-1). All models share the same architecture as the baseline, with subscript numbers indicating the relative improvement.}
\label{tab2}
\end{table}

\begin{table}[!ht]
  \renewcommand{\arraystretch}{1.8}
  \setlength{\tabcolsep}{3pt}
  \centering
  \scalebox{0.75}{
  \begin{tabular}{l c c c c c c}
    \hline
    \multirow{3}{*}{\textbf{Model}} & \multirow{3}{*}{\textbf{\makecell{Train\\Set\\ANR}}} & \multirow{3}{*}{\textbf{\makecell{Stream\\Decode}}} & \multicolumn{4}{c}{\textbf{Test WER($\%$) $\downarrow$}} \\
    \cline{4-7}
     & & & \textbf{Aishell-1 ANR} & \multicolumn{3}{c}{\textbf{WenetSpeech}} \\
    \cline{4-7}
     & & & \textbf{0$\%$} & \textbf{Dev} & \textbf{Test\_Net} & \textbf{Test\_Meeting} \\
    \hline
    \multirow{2}{*}{Baseline} & \multirow{2}{*}{0\%} & True & 5.33 & 29.90 & 48.94 & 53.99 \\
     & & False & \textbf{4.93} & 29.16 & 48.29 & 53.60 \\
    \hline
    \multirow{2}{*}{DGSNA\_10} & \multirow{2}{*}{10\%} & True & ${5.48}_{\color{Firebrick4}{+2.81\%}}$ & ${29.83}_{\color{Cyan4}{-0.23\%}}$ & ${47.45}_{\color{Cyan4}{-3.04\%}}$ & ${49.80}_{\color{Cyan4}{-7.76\%}}$ \\
     & & False & ${4.94}_{\color{Firebrick4}{+0.20\%}}$ & $\mathbf{{29.08}_{\color{Cyan4}{-0.27\%}}}$ & ${46.74}_{\color{Cyan4}{-3.21\%}}$ & ${49.36}_{\color{Cyan4}{-7.91\%}}$ \\
    \hline
    \multirow{2}{*}{DGSNA\_20} & \multirow{2}{*}{20\%} & True & ${5.52}_{\color{Firebrick4}{+3.56\%}}$ & ${29.84}_{\color{Firebrick4}{+0.20\%}}$ & ${46.98}_{\color{Cyan4}{-4.00\%}}$ & ${48.94}_{\color{Cyan4}{-9.35\%}}$ \\
     & & \cellcolor{LightCyan1}False & \cellcolor{LightCyan1}${4.96}_{\color{Firebrick4}{+0.61\%}}$ & \cellcolor{LightCyan1}${29.18}_{\color{Firebrick4}{+0.07\%}}$ & \cellcolor{LightCyan1}$\mathbf{{46.21}_{\color{Cyan4}{-4.31\%}}}$ & \cellcolor{LightCyan1}${48.42}_{\color{Cyan4}{-9.66\%}}$ \\
    \hline
    \multirow{2}{*}{DGSNA\_30} & \multirow{2}{*}{30\%} & True & ${5.69}_{\color{Firebrick4}{+6.75\%}}$ & ${30.28}_{\color{Firebrick4}{+1.27\%}}$ & ${47.87}_{\color{Cyan4}{-2.19\%}}$ & ${47.95}_{\color{Cyan4}{-11.19\%}}$ \\
     & & False & ${5.14}_{\color{Firebrick4}{+4.26\%}}$ & ${29.44}_{\color{Firebrick4}{+0.95\%}}$ & ${47.15}_{\color{Cyan4}{-2.36\%}}$ & $\mathbf{{47.53}_{\color{Cyan4}{-11.32\%}}}$ \\
    \hline
    \multirow{2}{*}{DGSNA\_40} & \multirow{2}{*}{40\%} & True & ${6.01}_{\color{Firebrick4}{+12.76\%}}$ & ${31.45}_{\color{Firebrick4}{+5.34\%}}$ & ${50.01}_{\color{Firebrick4}{+2.19\%}}$ & ${50.38}_{\color{Cyan4}{-6.69\%}}$ \\
     & & False & ${5.39}_{\color{Firebrick4}{+9.33\%}}$ & ${30.56}_{\color{Firebrick4}{+4.80\%}}$ & ${49.30}_{\color{Firebrick4}{+2.09\%}}$ & ${49.92}_{\color{Cyan4}{-6.87\%}}$ \\
    \hline
  \end{tabular}
  }
  \caption{Results of the ASR experiment (WenetSpeech). All models share the same architecture as the baseline, with subscript numbers indicating the relative improvement.}
\label{tab3}
\end{table}

\begin{figure}[!ht]
    \centering
    \includegraphics[width=\textwidth]{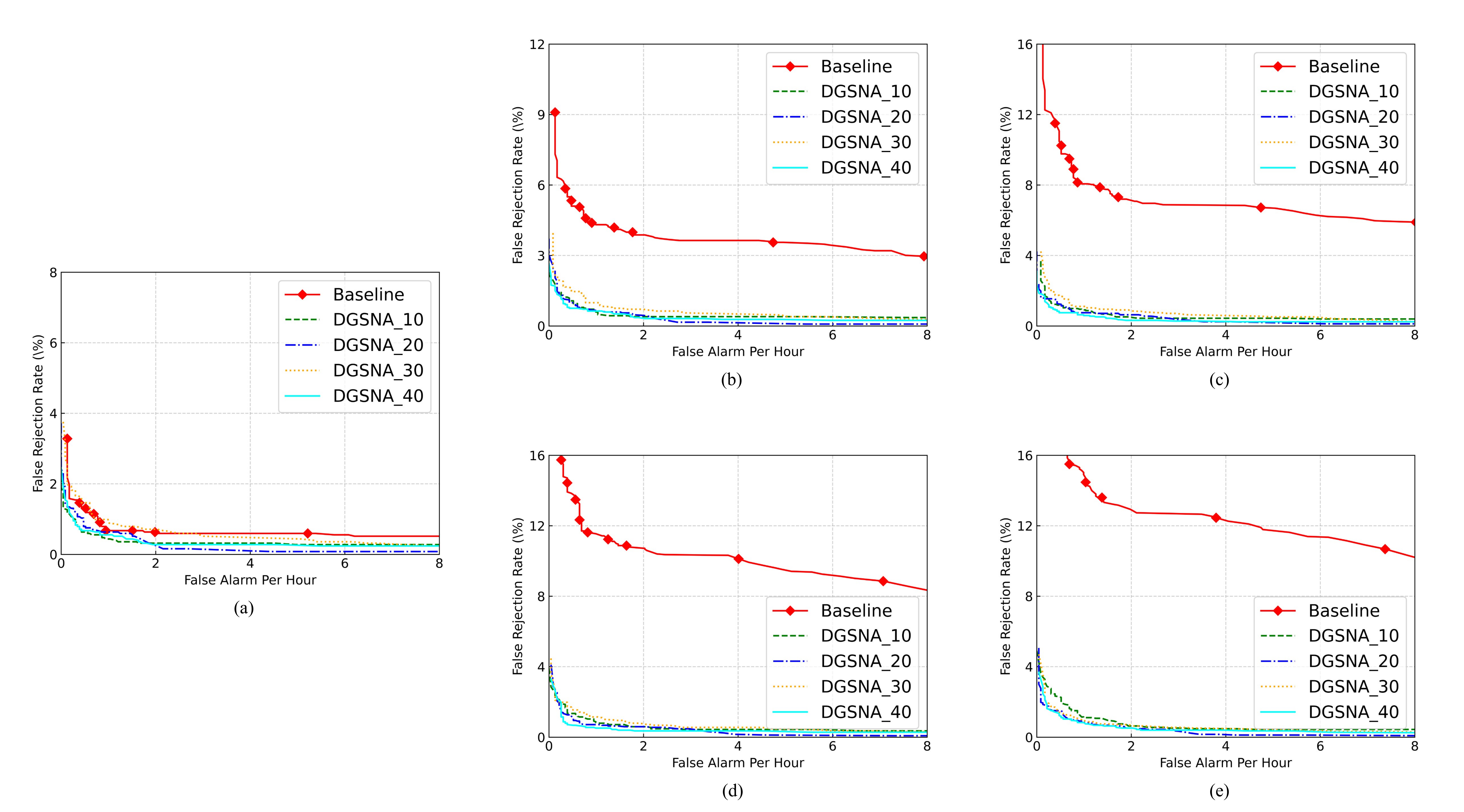}
    \caption{Results of the KWS experiment. All models share the same architecture as the baseline. The five subplots, presented alphabetically, show performance on the test set at ANRs of 0\%, 10\%, 20\%, 30\%, and 40\%.}\label{fig6}
\end{figure}

Tables \ref{tab2} and \ref{tab3} present the ASR experimental results. Introducing DGSNA during training significantly improves performance in both known noisy scenarios (Aishell-1 with ANR $>$ 0) and, to a lesser extent, unknown scenarios (WenetSpeech), achieving up to 11.32\% relative improvement. This improvement comes at the cost of slight performance degradation in clean scenarios (Aishell-1 with ANR = 0). The DGSNA application rate (ANR) significantly influences results. We hypothesize that a low ANR (10\%) results in insufficient data augmentation, while a high ANR (40\%) leads to overfitting. An ANR of 20\% proved optimal.

Figure \ref{fig6} presents the KWS experimental results. Introducing DGSNA during training significantly improves performance in known noisy scenarios, with an ANR of 20\%, once again demonstrating optimal effectiveness.

\subsection{Comparison of different generative chat models}
\label{subsec5_4}

\subsubsection{BET prompt}

The effectiveness of a generative LLM model in dynamic scene simulation significantly hinges on the adaptability of the BET prompts to the model's specific input-output structure.

\begin{figure}[ht]
    \centering
    \includegraphics[width=.96\textwidth]{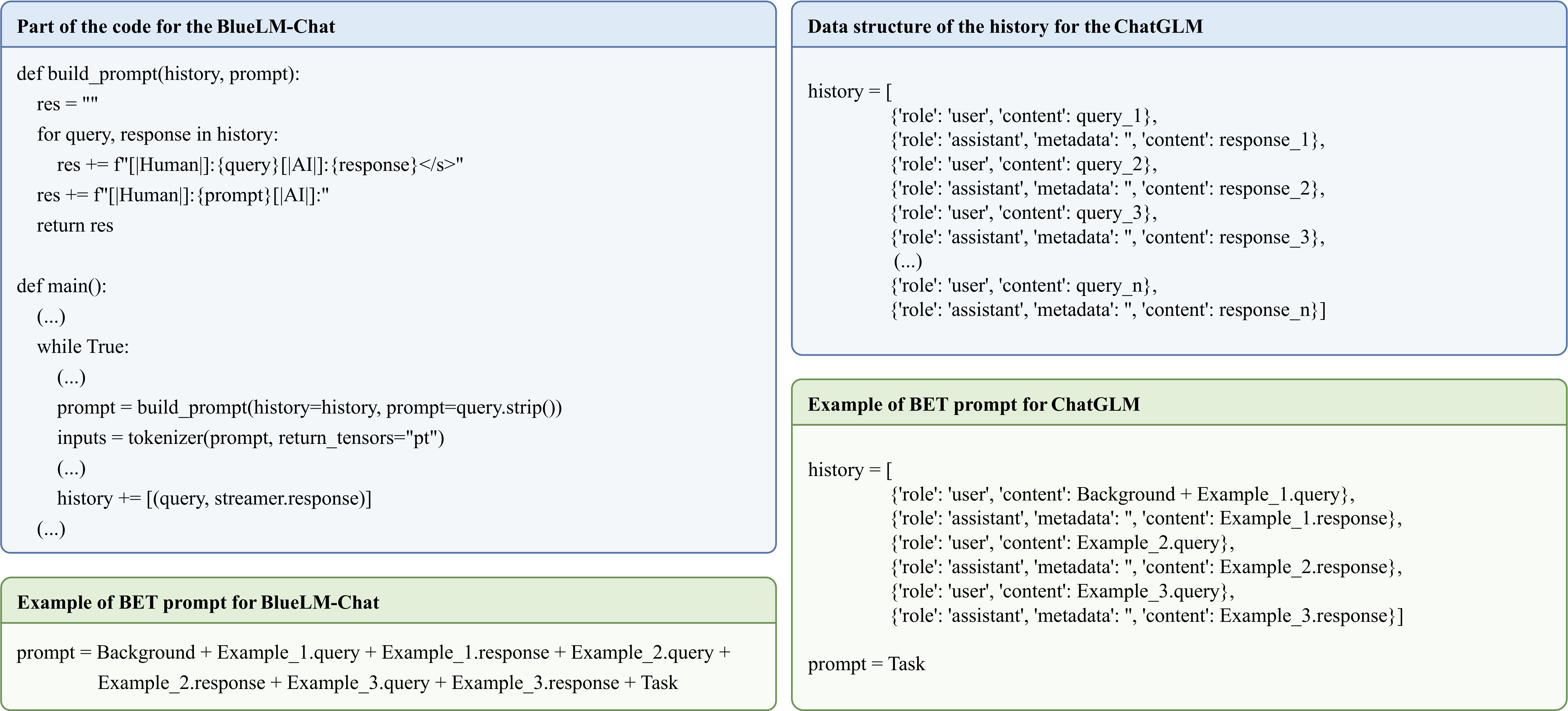}
    \caption{Construction of the BET prompt.}\label{fig7}
\end{figure}\textbf{}

To ensure compatibility across diverse model architectures, tailoring the BET prompt to the operational mode of the target generative LLM model is crucial. Generative chat models are commonly classified into two types based on their parameter-passing methods: (a) Single Parameter Input Models: In this configuration, models such as BlueLM-Chat concatenate the current chat input and historical chat information into a single consolidated parameter, typically termed the prompt. (b) Dual Parameter Input Models: Conversely, models like ChatGLM are designed to process the current chat input and historical chat information as two separate parameters, referred to as the prompt and history. As illustrated in Figure \ref{fig7}, we developed tailored prompts for two distinct models, BlueLM-Chat and ChatGLM, corresponding to their respective single and dual parameter input frameworks.

For single-parameter input models like BlueLM-Chat, where the prompt requires the integration of the current chat input with historical chat information into a single parameter, the construction of the BET prompt is sequential and cumulative. Each chat interaction builds upon the previous one. The BET prompt is constructed in the order of Background, Examples, and Task. For each Example, the prompt is further refined by concatenating the query and the response.

For dual parameter input models like ChatGLM, where the current chat input and the historical chat information are fed as separate parameters. Specifically, history is formatted as a sequence that includes the background, followed by alternating queries and responses from the examples. In contrast to the history, the prompt is straightforward and focused, being comprised solely of the Task. In conclusion, the construction of a BET prompt, as described, can be finely adjusted based on the specific input mode of various generative chat models.

\subsubsection{Filter metrics}

\begin{figure}[ht]
\centering
\includegraphics[width=.9999\textwidth]{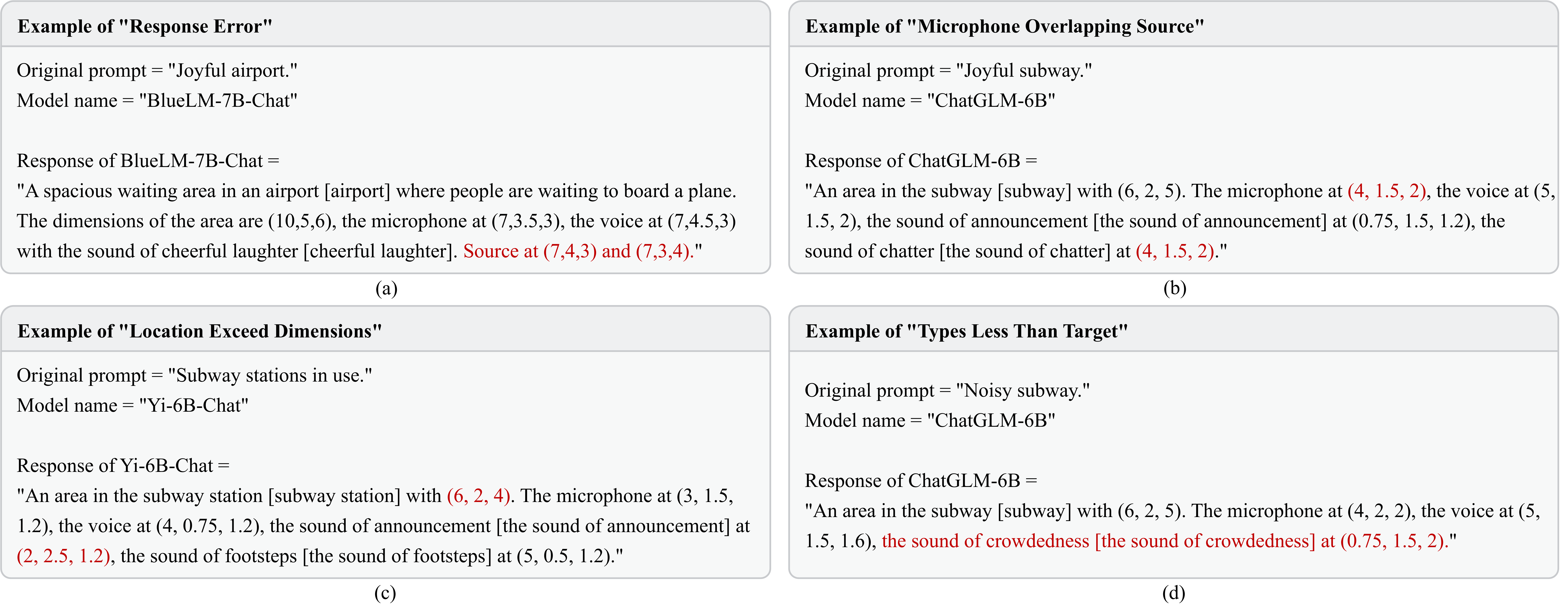}
\caption{Examples of each filter metric.}\label{fig8}
\end{figure}

To ensure that the generated scenes are suitable for subsequent scene-based noise addition, our research group developed and implemented a set of filter metrics. These metrics are specifically designed to assess the quality and applicability of responses generated by different generative chat models, effectively screening out those that do not meet established criteria. Figure \ref{fig8} provides illustrative examples for each metric, with red font denoting incorrect content. The specifics of these filter metrics are:

\begin{itemize}
\item \textbf{Response Error:} Verifies adherence to the 3-shot example format and the presence of essential information, such as scene dimensions and coordinates.
\item \textbf{Microphone Overlapping Source:} Ensures the microphone's position does not overlap any sound sources, preventing one sound from completely masking others.
\item \textbf{Location Exceed Dimensions:} Checks if locations in the response are within the defined scene dimensions.
\item \textbf{Types Less Than Target:} Measures whether the number of noise types specified in the response meets or exceeds a predefined target, set to 2 for this experiment.
\end{itemize}

Specifically, only comments that meet criterion 1 are evaluated by criteria 2, 3, and 4.

Table \ref{tab4} displays the pass rates of model replies under each prompt framework, with the BT framework providing only background (B) and task (T) and the ET framework providing only example (E) and task (T), to examine the effects of various prompt frameworks on DGSNA. The BET framework performed the best in the testing. This is due to the fact that whereas measure 1 can only be assessed by consulting the instances (E), metrics 2, 3, and 4 can be assessed by looking at the examples (E) or by following the background (B). Consequently, the model cannot offer an accurate response when the number of examples (E) is 0.

\begin{table}[ht]
  \renewcommand{\arraystretch}{1.6}
  \setlength{\tabcolsep}{3pt}
  \centering
  \scalebox{0.9}{
    \begin{tabular}{c c c c c c}
        \hline
        \multirow{2}{*}{framework} & \multirow{2}{*}{model} & \multicolumn{4}{c}{pass rates (\%) $\uparrow$} \\
        \cline{3-6}
            & & \#1 & \#2 & \#3 & \#4 \\
        \hline
        \multirow{2}{*}{BT} & ChatGLM & 0 & / & / & / \\
            & Qwen & 0 & / & / & / \\
        \hline
        \multirow{2}{*}{ET} & ChatGLM & 55.8 & 83.8 & 87.3 & 93.7 \\
            & Qwen & \textbf{98.1} & 93.0 & 82.2 & 97.2 \\
        \hline
        \multirow{2}{*}{BET} & ChatGLM & 56.7 & 84.3 & \textbf{96.1} & \textbf{100} \\
            & Qwen & 97.8 & \textbf{93.2} & 87.5 & \textbf{100} \\
        \hline
    \end{tabular}
  }
  \caption{Performance of four generative chat models against filter metrics.}\label{tab4}
  \end{table}

\subsection{Example analysis}
\label{subsec5_5}

By leveraging the BET prompt framework alongside generative LLM models, TTA models, and RIR filters, we can effectively implement the DGSNA. This section elucidates the specific process through which scene-based speech is generated, illustrated by the analysis of examples.

\subsubsection{Correct Example}

\begin{figure}[ht]
\centering
\includegraphics[width=.9999\textwidth]{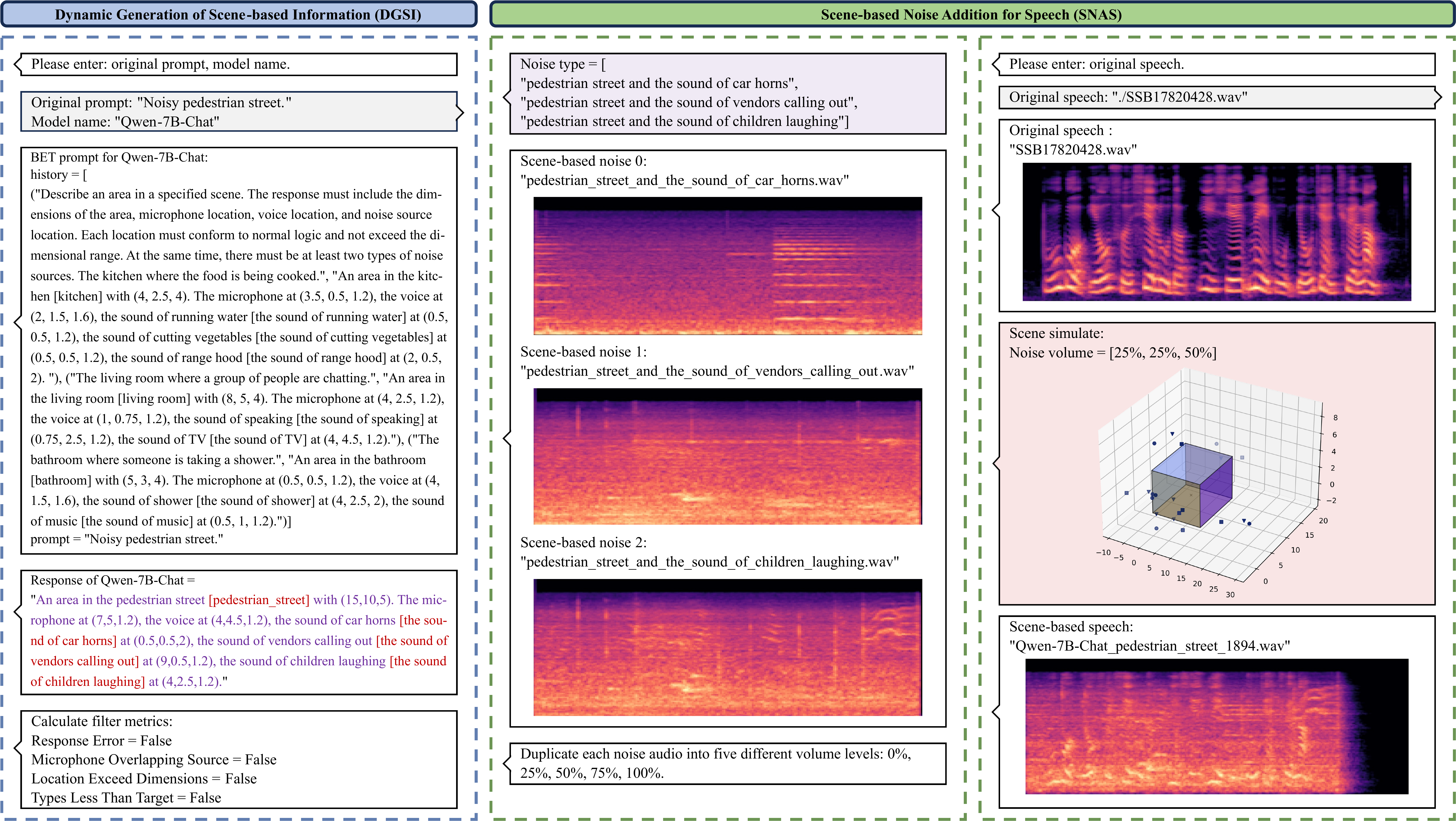}
\caption{Workflow for generating scene-based speech.}\label{fig9}
\end{figure}

This subsection provides a detailed analysis of the workflow for generating scene-based speech, using the sample depicted in Figure \ref{fig9} as an illustration. This process underscores the comprehensive steps undertaken to ensure the production of high-quality, contextually appropriate scene-based speech utilizing DGSNA.

\begin{itemize} 
\item \textbf{Original Prompt:} The process begins with the input text ``noisy pedestrian street."
\item \textbf{BET Prompt:} The BET prompt has been strategically adapted to conform to the input-output structure and historical context management strategy employed by the Qwen-7B-Chat model.
\item \textbf{Scene-based Information:} The adapted BET prompt is fed into the Qwen-7B-Chat model, which processes the information and dynamically generates scene-based information.
\item \textbf{Filter Metrics:} The generated scene-based information is assessed using predefined filter metrics to confirm its suitability for subsequent processing.
\item \textbf{TTA model:} The TTA model utilizes each noise type specified in the scene-based information as a text embedding. These embeddings undergo a denoising procedure to generate precise audio prior representations. Subsequently, these representations are converted into actual audio samples through the combined use of a decoder and a vocoder.
\item \textbf{Scene-based Noise:} Each audio sample is generated at five distinct volume levels, mimicking real-world variations in sound intensity. To improve the realism of scene-based noise addition and more accurately simulate real-world acoustic environments, our research group incorporated a parameter for noise source volume levels. Employing torchaudio \citep{ref62}, five distinct volume levels—0\%, 25\%, 50\%, 75\%, and 100\%—were generated for each noise audio sample. During simulation, a random selection process determined the active volume level for each noise source.
\item \textbf{Original Speech:} The original speech is obtained.
\item \textbf{Scene-based Speech:} The RIR filter uses the scene dimensions and source coordinates provided in the scene-based information to construct the RIR. Subsequently, it convolves both the generated scene-based noise and original speech with the corresponding RIR, culminating in the production of the scene-based speech.
\end{itemize}

We present the scene visualization and the example speech's spectrogram in Figure \ref{fig9}. Squares, inverted triangles, and circles each denote a unique type of sound source and its mirror sources. The number of mirror sources (e.g., 18) is calculated based on the product of the number of original sound sources (e.g., 3), the number of scene surfaces (e.g., 6), and the maximum specified order of reflection (e.g., 1). In outdoor scenes, these mirror sources are treated as additional sound sources of the same type. By including both original and mirror sources, the total number of sound sources in the scene amounts to 21. This comprehensive inclusion ensures a more detailed and accurate simulation, as it accounts for multiple interactions and the resultant acoustic effects within the scene.

\subsubsection{Incorrect Example}

\begin{figure}[ht]
  \centering
  \includegraphics[width=.9999\textwidth]{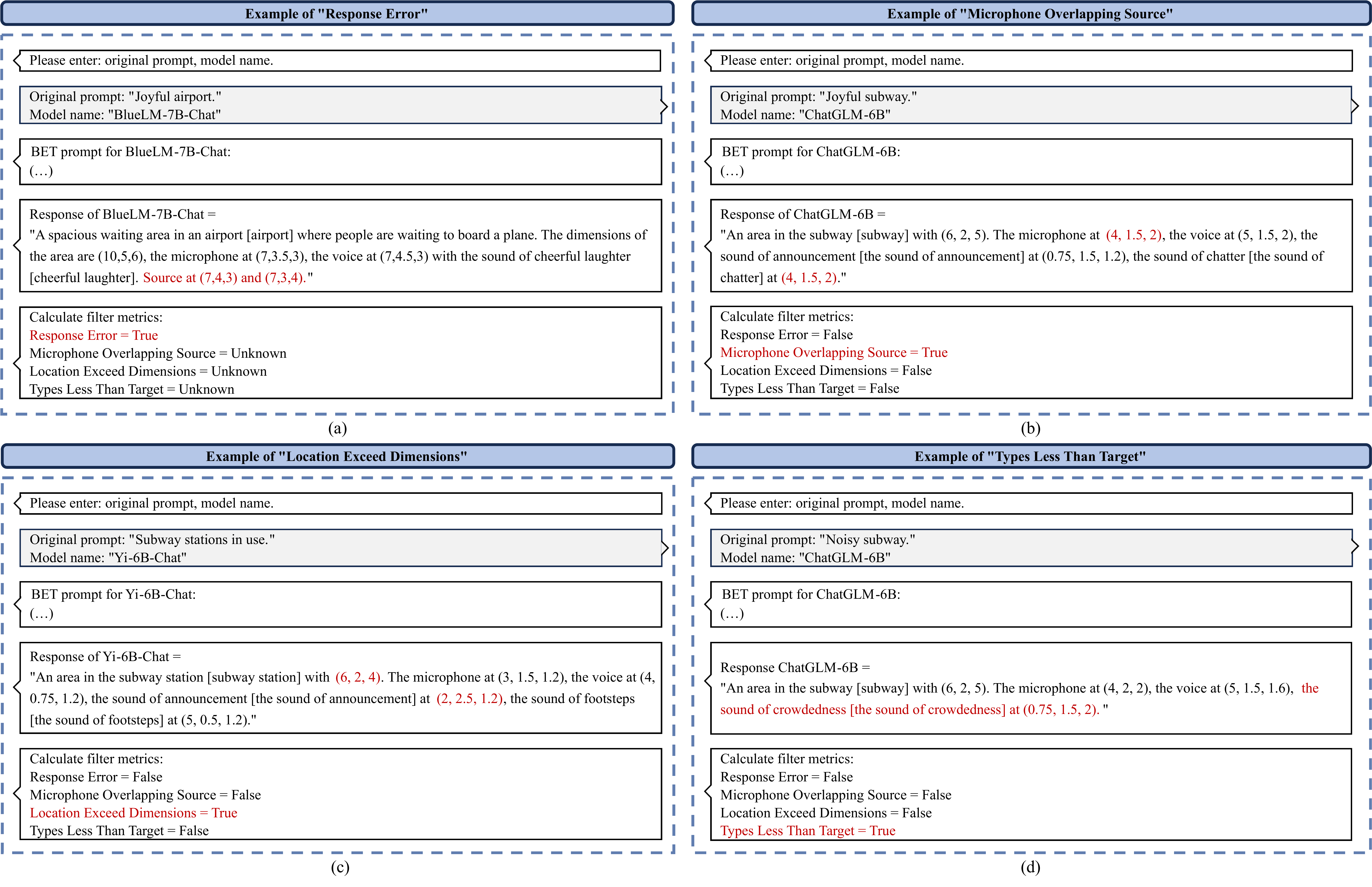}
  \caption{Workflow for generating four specific examples.}\label{fig10}
\end{figure}

In this section, the analysis of four specific examples from Section \ref{subsec5_3} is detailed, each illustrating a distinct issue encountered in the application of the DGSNA. Each example is associated with a particular filter metric that has been triggered due to identified errors, demonstrating how the system can prevent incorrect or inadequate data from progressing through the audio generation process. Figure \ref{fig10} outlines the general workflow for four specific examples. Here is the analysis of specific issues:

\begin{itemize} 
    \item \textbf{First Example (Figure \ref{fig10}.a):} The response fails to adhere to the 3-shot format provided, contains logical inconsistencies, lacks essential details such as scene dimensions and coordinate points, and fails to provide translations.
    \item \textbf{Second Example (Figure \ref{fig10}.b):} Positioning the microphone to overlap the source of the crowd chatter helps prevent the background noise from completely masking other sounds.
    \item \textbf{Third Example (Figure \ref{fig10}.c):} Locations mentioned in the response exceed the defined dimensions of the scene. Such discrepancies can lead to computational errors (e.g., ValueError) when the data is used in subsequent processes like the RIR filter.
    \item \textbf{Fourth Example (Figure \ref{fig10}.d):} The response includes fewer noise types than the targeted number set in the BET prompt, which is specified as 2. Insufficient variety in noise types compromises the depth and authenticity of the scene simulation, reducing its effectiveness in creating a believable acoustic environment.
\end{itemize}

\section{Conclusion}
\label{sec6}
This paper aims to address the limitations of existing speech noise addition methods, such as the limited coverage of real-world acoustic scenes, heavy dependence on pre-recorded noise libraries, and static scene metadata. 
We introduce Dynamic Generative Scene-based Noise Addition (DGSNA), a novel framework that integrates DGSI and SNAS modules. Our approach leverages generative LLMs and a TFD-based TTA model to create diverse, context-aware noise environments.

The DGSI module adopts a self-designed BET prompt framework to guide generative LLMs in dynamically generating logically consistent scene information and is compatible with LLM architectures via tailored prompt construction; filter metrics are also designed to ensure the validity of generated scene information. The SNAS module constructs a TFD-based TTA model with a TFC Audio Encoder to synthesize scene-specific noise and fuses the noise with clean speech by RIR filters based on the ISM, thus streamlining the acoustic environment simulation process.

Extensive experiments on ASR and KWS tasks verify the effectiveness and superiority of DGSNA. The method significantly improves the robustness of downstream speech models, achieving a relative performance improvement of up to 11.32\% in unknown acoustic scenes, with the optimal ANR of 20\%. DGSNA shows high compatibility with existing techniques like SpecAugment, and their combination yields better performance. 

By moving beyond the need for exhaustive real-world acoustic catalogs, this study addresses traditional data bottlenecks and establishes a new generative framework for speech data augmentation.
The BET prompt framework and TFD-based TTA model also offer new insights into the integration of generative AI and speech signal processing. For future research, we will focus on optimizing DGSNA for complex fused acoustic scene generation, long-duration noise synthesis, and high-precision RIR simulation of irregular spaces and further validate its generalization performance on low-resource and multilingual speech datasets, as well as its applicability to more speech tasks such as speech separation and speaker verification.

\end{document}